\def\tr{{\text{tr}}\,}
\def\Tr{{\text{Tr}}\,}
\def\Im{{\text{Im}}\,}
\def\Re{{\text{Re}}\,}
\def\kF{k_{\text{F}}}
\def\vF{v_{\text{F}}}
\def\NF{N_{\text{F}}}
\def\me{m_{\text{e}}}

\def\sgn{{\text{sgn\,}}}
\def\qslash{{q\hskip -5pt /}}
\def\be{\begin{equation}}
\def\ee{\end{equation}}
\def\bea{\begin{eqnarray}}
\def\eea{\end{eqnarray}}
\def\bse{\begin{subequations}}
\def\ese{\end{subequations}}

\def\qslash{q\!\!\!/}
\documentclass[prb,twocolumn,showpacs,amsmath,amssymb,eqsecnum,preprintnumbers]{revtex4}
\usepackage{graphicx} 
\usepackage{dcolumn} 
\usepackage{bm} 
\usepackage{amsfonts}
\newcommand{\field}[1]{\mathbb{#1}}
\newcommand{\C}{\field{C}} 
\newcommand{\R}{\field{R}} 
\newcommand{\fourchoose}[5]{\left(\begin{array}{c}\genfrac{}{}{0pt}{}{#1}{#2}\\[0.4em]
       \genfrac{}{}{0pt}{}{#3}{#4}\end{array}\right)_{#5}} 
\draft
\begin{document}
\title{Effective Soft-Mode Theory of Strongly Interacting Fermions}
\author{D. Belitz$^{1,2}$, T. R. Kirkpatrick$^3$}
\affiliation{$^{1}$ Department of Physics and Institute of
Theoretical Science,
University of Oregon, Eugene, OR 97403, USA\\$^{2}$ Materials Science
Institute, University of Oregon,
Eugene,
                    OR 97403, USA\\
$^{3}$ Institute for Physical Science and Technology,and Department of
Physics, University of Maryland, College Park, MD 20742, USA\\
}
\date{\today}

\begin{abstract}
An effective field theory for clean electron systems is developed in analogy to the generalized
nonlinear sigma model for disordered interacting electrons. The physical goal is to separate
the soft or massless electronic degrees of freedom from the massive ones and integrate out
the latter to obtain a field theory in terms of the soft degrees of freedom only. The resulting 
theory is not perturbative with respect to the electron-electron interaction. It is controlled by 
means of a systematic loop expansion and allows for a renormalization-group analysis in a 
natural way. It is applicable to universal phenomena within phases, and to transitions between 
phases, with order parameters in arbitrary angular-momentum channels, and in the 
spin-singlet, spin-triplet, particle-hole, and particle-particle channels. Applications
include ferromagnetic and ferrimagnetic ordering, non-s-wave ferromagnetic order 
(magnetic nematics), Fermi-liquid to non-Fermi-liquid
transitions, and universal phenomena within a Fermi-liquid phase. 
\end{abstract}
\pacs{71.10.-w; 71.10.Pm; 71.10.Hf; 71.10.+a}
\maketitle

\section{Introduction}
\label{sec:I}

The theory of many-fermion systems is of obvious importance because of its ubiquitous applications in condensed-matter physics.
It is also a very hard problem, due to the difficulty of dealing with a macroscopic number of electrons which, in many materials,
interact strongly enough to preclude treating the interactions as weak perturbations of noninteracting electrons. Historically, there 
have been two main approaches to the problem. Landau's Fermi-liquid theory,\cite{Lifshitz_Pitaevskii_1980, Baym_Pethick_1991}  
although not perturbative, is phenomenological and self-contained in nature, and often a more microscopic approach is desirable, 
especially for investigations of the limitation of Fermi-liquid theory. Many-body diagrammatic 
techniques,\cite{Abrikosov_Gorkov_Dzyaloshinski_1963, Fetter_Walecka_1971} on the other hand, are limited by their inherently 
perturbative approach. Furthermore, it is neither feasible nor desirable to study all aspects of a microscopic model, except in some special cases. For many important 
applications it suffices to keep only the low-lying excitations or soft modes of the system, which govern the behavior at long times and large 
distances. The most obvious examples of  such applications are phase transitions, but there also are many qualitative properties 
of entire phases that depend on the soft modes only. These observables also have the attractive property of being universal in the 
sense that they depend only on basic symmetry properties of the system, and not on any microscopic details. Consider, for instance, 
a Heisenberg ferromagnet. The critical behavior at the Curie point, i.e., the phase transition from the paramagnetic to the 
ferromagnetic phase, is entirely governed by soft modes,\cite{Ma_1976} but so is the divergence of the longitudinal susceptibility everywhere in 
the ordered phase.\cite{Brezin_Wallace_1973} To describe and understand either phenomenon it therefore suffices to consider an effective theory that takes 
into account only the relevant soft modes. In the case of the Heisenberg ferromagnet at temperatures $T>0$, either a classical 
$O(3)$ $\phi^4$-theory or an $O(3)$ nonlinear sigma model provides a suitable effective theory.\cite{Zinn-Justin_1996} Such a theory can sometimes
be constructed based on symmetry considerations only, but it often is desirable, and leads to greater physical insight, to start with 
a microscopic description and derive the effective theory by integrating out all non-essential degrees of freedom in some simple
approximation that respects the crucial symmetries. The resulting ``hydrodynamic'', i.e., long-wavelength and low-frequency,
description of any system of interest is complementary to, and serves a very different purpose than, for instance, a first-principles
approach that aims at a description at atomic length and energy scales. In order to obtain a well-behaved effective theory, care
must be taken to keep all soft modes of interest {\em and} all other soft modes that couple to them. Integrating out soft modes leads
to long-ranged effective interactions between the remaining degrees of freedom, and hence to non-local effective field theories
that can be hard to handle technically.

For noninteracting disordered electrons such an effective-theory approach was pioneered by 
Wegner.\cite{Wegner_1979} Guided by an analogy with classical ferromagnets he constructed 
a matrix nonlinear sigma model capable of describing the instability of a
(disordered) Fermi-liquid phase against the formation of an Anderson insulator, and allowed for 
the determination of the critical behavior at the Anderson transition in terms of an 
$\epsilon$-expansion about the lower critical dimension for this problem,
$d_{\text c}^- = 2$. Wegner's original formulation was expanded upon by various 
authors,\cite{Schaefer_Wegner_1980, Pruisken_Schaefer_1982, Efetov_Larkin_Khmelnitskii_1980} 
and eventually his sigma model was generalized to interacting disordered 
electrons.\cite{Finkelstein_1983} This theory and its generalizations have been extensively used to
describe the Anderson-Mott metal-insulator transition.\cite{Finkelstein_1984a, Belitz_Kirkpatrick_1994}
These theories are fundamentally based on diffusive soft modes, and do not allow for the clean limit to
be taken. This led to the strange situation where a successful effective soft-mode theory
was available for disordered electrons, but not for clean ones, although one might think that the 
latter poses a simpler problem.\cite{Griffiths_footnote} This last conclusion is fallacious, however, 
mostly because the number of soft modes is much larger in clean electron systems
than in disordered ones. This was noted in Ref.\ \onlinecite{Belitz_Kirkpatrick_1997}, whose main
objective was to provide a thorough derivation of Finkelstein's generalized nonlinear sigma model 
for disordered systems. While this theory did allow for the clean limit to be taken, the resulting field
theory had non-local vertices and did not offer any obvious way to obtain a description as useful 
as the nonlinear sigma model for the disordered case. Another effective field
theory for clean electrons was developed in Ref.\ \onlinecite{Aleiner_Efetov_2006}. This theory 
is technically based on a supersymmetric matrix formulation and aims to incorporate and unify 
various other approaches that discuss the validity and limitations of Landau Fermi-liquid 
theory.\cite{Chitov_Senechal_1998, Metzner_Castellani_DiCastro_1998, Chubukov_Maslov_2003, 
Chubukov_Maslov_2004, Chubukov_et_al_2005, Chubukov_Maslov_2007} It introduces bosonic 
degrees of freedom
and integrates out the fermions, as did Ref.\ \onlinecite{Belitz_Kirkpatrick_1997}, in contrast 
to Shankar's derivation of Fermi-liquid theory that applied renormalization-group (RG) techniques 
to a fermionic field theory.\cite{Shankar_1994} The theory of Ref.\ \onlinecite{Aleiner_Efetov_2006}
has recently been refined\cite{Efetov_Pepin_Meier_2009} and formulated both in a version that is
formally exact and may be better suitable for numerical work than standard Monte-Carlo 
techniques,\cite{Efetov_Pepin_Meier_2010} and as an effective field theory for low-lying
excitations.\cite{Meier_Pepin_Efetov_2011} The latter was used for an analysis of the
low-temperature specific heat in two-dimensional systems. 

In the present paper we take a different approach to this problem. It is similar in spirit to
Refs.\ \onlinecite{Aleiner_Efetov_2006, Meier_Pepin_Efetov_2011}, but rather different
technically. Our physical motivation is the
desirability of a broadly useful effective theory that is capable of describing widely different 
phenomena such as ferromagnetic ordering in non-s-wave angular momentum 
channels,\cite{Oganesyan_Kivelson_Fradkin_2001, Kirkpatrick_Belitz_2011a} quantum ferrimagnetic
order, which has received little attention so far, 
a possible breakdown of Fermi-liquid theory due to a vanishing quasi-particle 
weight,\cite{Kirkpatrick_Belitz_2011b}
and, in addition, novel universal aspects of the Fermi-liquid phase.\cite{Kirkpatrick_Belitz_2011b} We are interested in a local
field theory that is not perturbative with respect to the electron-electron interaction, but rather can be controlled by means
of a systematic loop expansion. This will provide a basis for an application of RG techniques that 
allow for a resummation of the expansion in powers of the loop parameter, and thus allow to go beyond perturbation theory
in a controlled fashion. This not possible within the framework of traditional perturbation 
theory,\cite{Abrikosov_Gorkov_Dzyaloshinski_1963, Fetter_Walecka_1971} which provides no information about
the structure of the renormalized theory. Technically, the theory we will develop is based on a generalization of 
Ref.\ \onlinecite{Belitz_Kirkpatrick_1997}. The reason for the non-locality of the latter was its restriction to density or
s-wave ($\ell=0$) modes, with soft modes in higher angular-momentum channels being effectively integrated out. Formulating the
theory in terms of phase-space variables allows one to keep the soft modes in all angular-momentum channels. This
leads both to a local theory and allows for natural description of order in $\ell\neq 0$ channels. Integrating out the massive
modes in a tree approximation leads to an effective action that explicitly keeps all of the soft modes. Its structure is
different from that of the generalized nonlinear sigma model for the disordered case. It can be constructed to any 
desired order in the soft degrees of freedom, which in turn allows to perform a loop expansion to any desired order.

The organization of this paper is as follows. In Sec.\ \ref{sec:II} we formulate our model in terms of fermionic fields. 
We then bosonize the theory by constraining bilinear products of fermion fields to bosonic matrix fields $Q$ and
integrating out the fermions. We then discuss symmetry properties of the $Q$-fields and representations of
observables in terms of $Q$-correlation functions. In Sec.\ \ref{sec:III} we identify the soft modes of the action by
means of a Ward identity. In Sec. \ref{sec:IV} we expand the $Q$-field theory about a saddle-point solution that
describes a Fermi liquid, and separate the soft fluctuations from the massive ones by means of the Ward identity.
The massive degrees of freedom are then integrated out in a tree approximation to derive an effective theory
entirely in terms of the soft modes. We also make contact with the density formulation of Ref.\ \onlinecite{Belitz_Kirkpatrick_1997}.
In Sec.\ \ref{sec:V} we discuss universal features of the energy-dependent density of states
(DOS) in a Fermi-liquid as a simple application.
In Sec.\ \ref{sec:VI} we summarize and discuss the effective theory. We finally discuss various
future applications of the theory. A discussion of the effects of quenched disorder, and a pedagogical discussion
of the $O(2)$ nonlinear sigma model that stresses analogies with the current theory, are relegated to two appendixes.

\section{Field-theoretic formulation of the interacting fermion
problem}
\label{sec:II}

\subsection{Fermionic formulation}
\label{subsec:II.A}
Our starting point is a description of interacting fermions in terms of Grassmann 
fields. The partition function can be written\cite{Negele_Orland_1988}
\be
Z = \int D[{\bar\psi},\psi]\ e^{S[{\bar\psi},\psi]}\quad.
\label{eq:2.1}
\ee
Here ${\bar\psi}$ and $\psi$ are Grassmann-valued fields , and $D[{\bar\psi},\psi]$ is 
the Grassmannian integration measure. The action $S$ has the form
\bse
\label{eqs:2.2}
\be
S = \int dx \sum_{\sigma}
{\bar\psi}_{\sigma}(x)\,\left[-\partial_{\tau} - \epsilon({\bm\nabla})+ \mu \right]\,
     \psi_{\sigma}(x) + S_{\text{int}}\ .
\label{eq:2.2a}
\ee
The first term describes noninteracting electrons with chemical potential $\mu$. 
For simplicity and definiteness we will assume a single parabolic band, i.e.,
\be
\epsilon({\bm\nabla}) = -{\bm\nabla}^2/2\me
\label{eq:2.2b}
\ee
with $\me$ the effective electron mass. If desired, the model can easily be
generalized to include a nontrivial band structure instead.
We use a $(d+1)$-vector space-time notation with $x = ({\bm x},\tau)$, 
$\int dx = \int_V d{\bm x}\, \int_0^{1/T} d\tau$. ${\bm x}$ denotes the position, 
$\tau$ is the imaginary-time variable, $V$ and $T$ are the system volume and the 
temperature, respectively, and $\sigma = \uparrow,\downarrow \equiv +,-$ is the spin label. 
We use units such that 
$\hbar = k_{\text B} = 1$. $S_{\text{int}}$ describes an electron-electron interaction 
via a two-body potential $v({\bm x})$,
\bea
S_{\text{int}} &=& -\frac{1}{2} \int dx_1\,dx_2
\sum_{\sigma_1,\sigma_2} v({\bm x}_1 - {\bm x}_2)\,
\delta(\tau_1 - \tau_2)
\nonumber\\
&&\hskip 0pt\times{\bar\psi}_{\sigma_1}(x_1)\,{\bar\psi}_{\sigma_2}(x_2)\,
                  \psi_{\sigma_2}(x_2)\,\psi_{\sigma_1}(x_1)\ .
\label{eq:2.2c}
\eea
\ese

In a microscopic theory the potential $v({\bm x})$ would be the Coulomb
interaction. Here we assume for simplicity that the theory has already been
renormalized to take into account screening, so $v({\bm x})$ represents a statically
screened Coulomb interaction, or a some similar short-ranged model interaction.
Equation (\ref{eq:2.2c}) describes the interactions of number-density fluctuations
at all wavelengths, and in a microscopic description this is the only interaction
term there is.\cite{nonrelativistic_footnote} Integrating out fluctuations at short
wavelengths in order to generate an effective long-wavelength theory generates
interaction amplitudes in the spin-triplet channel in addition to the spin-singlet
one, and in the particle-particle channel in addition to the particle-hole hole
one, and all of these effective interaction amplitudes appear in all angular
momentum channels.\cite{AGD_Sec_18, Belitz_Kirkpatrick_1997, 
Belitz_Kirkpatrick_2010b} The theory we will develop
is general, and any desired interaction amplitudes can be kept. However, for
the sake of transparency and simplicity of the formalism we will restrict 
ourselves to a finite number. Some of the most interesting applications of
the theory are related to instabilities of the Fermi-liquid state due to a 
density-density interaction, and to magnetic order of the ferromagnetic 
or magnetic nematic variety. Furthermore, the soft modes in the
particle-particle channel become massive in the presence of a magnetic
field, as they do in disordered systems,\cite{Belitz_Kirkpatrick_1994}
and hence can be suppressed experimentally. Accordingly, in the remainder
of this paper we keep only interaction amplitudes in the particle-hole channel, 
and further restrict ourselves to the spin-singlet s-wave ($\ell=0$) and the spin-triplet
s-wave ($\ell=0$) and p-wave ($\ell=1$) channels. Generalizations to other
interaction channels, for instance, the spin-triplet d-wave channel that has
been discussed in the literature,\cite{Oganesyan_Kivelson_Fradkin_2001}
are straightforward. It needs to be stressed, however, that a restriction to
low-angular-momentum channels represents an approximation over and
above the low-energy restriction that is central to our approach, and that
a complete low-energy effective theory needs to keep all angular momentum
channels.

We define Fourier transforms of the fermionic field
\bea
{\bar\psi}_{\sigma}(k) &\equiv& {\bar\psi}_{n\sigma}({\bm k}) = \sqrt{T/V} \int dx\ e^{-ikx}\,{\bar\psi}_{\sigma}(x),
\nonumber\\
\psi_{\sigma}(k) &\equiv& \psi_{n\sigma}({\bm k}) = \sqrt{T/V} \int dx\ e^{ikx}\,\psi_{\sigma}(x),
\label{eq:2.3}
\eea
where $k = ({\bm k},\omega_n)$ is a four-vector that comprises a wave
vector ${\bm k}$ and a fermionic Matsubara frequency $\omega_n = 2\pi T(n+1/2)$,
and $kx \equiv {\bm k}\cdot{\bm x} - \omega_n\tau$. With the restrictions
explained above the action then reads
\bse
\label{eqs:2.4}
\be
S = \sum_{k,\sigma} {\bar\psi}_{\sigma}(k)\,\left[i\omega_n - {\bm k}^2/2\me
      + \mu\right]\, \psi_{\sigma}(k) + S_{\text{int}},
\label{eq:2.4a}
\ee
with
\bea
S_{\text{int}} &=& -\frac{\Gamma_{\text{s}}^{(0)}}{2}\, \frac{T}{V}  {\sum_q}^{\prime} \sum_{i=1}^3 n(q)\,n(-q) 
\nonumber\\
&&+\frac{\Gamma_{\text{t}}^{(0)}}{2}\, \frac{T}{V} {\sum_q}^{\prime} n_{\text{s}}^i(q)\,n_{\text{s}}^i(-q)
\nonumber\\
&& + \frac{\Gamma_{\text{t}}^{(1)}}{2}\, \frac{T}{V} {\sum_q}^{\prime} \sum_{i,\alpha=1}^{3}
       j_{\text{s}}^{i\alpha}(q)\,j_{\text{s}}^{i\alpha}(-q)
\label{eq:2.4b}
\eea
\ese
Here $q=({\bm q},\Omega_n)$ comprises a wave vector ${\bm q}$ and a bosonic Matsubara
frequency $\Omega_n = 2\pi T n$.
$\Gamma_{\text s}^{(0)}$, $\Gamma_{\text t}^{(0)}$, and $\Gamma_{\text s}^{(1)}$ are
the interaction amplitudes in the particle-hole spin-singlet s-wave, spin-triplet
s-wave, and spin-triplet p-wave channels, respectively. They are related to the parameters
$F_0^{\text s}$, $F_0^{\text a}$, and $F_1^{\text a}$, respectively, in Landau Fermi-liquid
theory.\cite{Abrikosov_Gorkov_Dzyaloshinski_1963} $\sum_{\bm q}^{\prime}$ denotes a 
sum over wave vectors that is restricted to $\vert{\bm q}\vert < \Lambda$ with a cutoff wave 
number $\Lambda$. The long-wavelength properties we are interested in do not depend on
$\Lambda$. The electron number density $n$, spin density $n_{\text s}$, and spin current
density $j_{\text s}^{i\alpha}$ are given in Eqs.\ (\ref{eqs:2.5}) below, and for completeness
we also list the number current density $j^{\alpha}$:
\bse
\label{eqs:2.5}
\bea
n(q) &=& \sum_k \sum_{\sigma} {\bar\psi}_{\sigma}(k+q/2)\,\psi_{\sigma}(k-q/2),
\label{eq:2.5a}\\
j^{\alpha}(q) &=& \sum_k k_{\alpha} \sum_{\sigma} {\bar\psi}_{\sigma}(k+q/2)\,\psi_{\sigma}(k-q/2),
\label{eq:2.5b}\\
n_{\text s}^i (q) &=& \sum_k \sum_{\sigma_1,\sigma_2} {\bar\psi}_{\sigma_1}(k+q/2)\,
                                    (\sigma_i)_{\sigma_1 \sigma_2}\,\psi_{\sigma_2}(k-q/2) ,
\nonumber\\
\label{eq:2.5c}\\
j_{\text s}^{i\alpha} (q) &=& \sum_k k_{\alpha}\! \sum_{\sigma_1,\sigma_2}
       {\bar\psi}_{\sigma_1}(k+q/2)
                                    (\sigma_i)_{\sigma_1 \sigma_2}\psi_{\sigma_2}(k-q/2) .
\nonumber\\
\label{eq:2.5d}
\eea
\ese
The $\sigma_i$ ($i=1,2,3$) are the Pauli matrices, and the $k_{\alpha}$
($\alpha=1,2,3$) are the components of the wave vector ${\bm k}$.
$\Gamma_{\text s}^{(0)} > 0$ describes a repulsive density-density interaction,
and $\Gamma_{\text t}^{0,1} > 0$ describe ferromagnetic interactions in the
s- and p-wave channels, respectively.

\subsection{Bosonic formulation}
\label{subsec:II.B}

\subsubsection{Phase-space degrees of freedom}
\label{subsubsec:II.B.1}

Our goal is to rewrite the action in terms of bosonic matrix fields. For this purpose
it is convenient to define a bispinor\cite{Efetov_Larkin_Khmelnitskii_1980}
\bse
\label{eqs:2.6}
\be
\eta_n({\bf x}) = 
     \frac{1}{\sqrt{2}}\left(
    \begin{array}{c}\ \ \bar{\psi}_{n\uparrow}({\bf x}) \\
                    \ \ \bar{\psi}_{n\downarrow}({\bf x})\\
                    \ \ \psi_{n\downarrow}({\bf x})\\ 
                    -\psi_{n\uparrow}({\bf x})\end{array}
                     \right)\quad,
\label{eq:2.6a}
\ee
with an adjoint\cite{adjoint_footnote}
\bea
\eta_n^+({\bm x}) &=&(C\eta)_n({\bm x})
\label{eq:2.6b}\\
                             &=& \frac{i}{\sqrt{2}}\left(-\psi_{n\uparrow}({\bm x}),-\psi_{n\downarrow}({\bm x}),
                                {\bar\psi}_{n\downarrow}({\bm x}),-{\bar\psi}_{n\uparrow}({\bm x})\right).
\nonumber
\eea
\ese
Here $C_{nm} = i(\tau_1\otimes s_2)\,\delta_{nm}$ is the charge conjugation matrix in the 
spin-quaternion space
spanned by $\tau_i\otimes s_j$ ($i,j = 0,1,2,3$) with $\tau_i = -s_j = -i\sigma_j$ with 
$\sigma_0$ the $2\times 2$ unit matrix and $\sigma_{1,2,3}$ the Pauli matrices.
Explicitly,
\begin{widetext}
\bse
\label{eqs:2.7}
\bea
\tau_0 &=& \left(\begin{array}{cc} 1 & 0 \\
                                                  0 & 1      \end{array}\right)\quad,\quad
\tau_1 = \left(\begin{array}{cc} 0 & -i \\
                                                  -i & 0      \end{array}\right)\quad,\quad
\tau_2 = \left(\begin{array}{cc} 0 & -1 \\
                                                  1 & 0      \end{array}\right)\,,\quad
\tau_3 = \left(\begin{array}{cc} -i & 0 \\
                                                  0 & i      \end{array}\right)\quad,\quad
\label{eq:2.7a}\\
s_0 &=& \left(\begin{array}{cc} 1 & 0 \\
                                                  0 & 1      \end{array}\right) \quad,\quad
s_1 = \left(\begin{array}{cc} 0 & i \\
                                             i & 0      \end{array}\right)\qquad\, ,\quad
s_2 = \left(\begin{array}{cc} 0 & 1 \\
                                                  -1 & 0      \end{array}\right)\,,\quad
s_3 = \left(\begin{array}{cc} i & 0 \\
                                                  0 & -i      \end{array}\right)\quad.
\label{eq:2.7b}
\eea
\ese
We now define a bilinear tensor product
\bea
B_{nm}({\bm x},{\bm y}) &=& \eta_n^+({\bm x})\otimes\eta_m({\bm y})
\nonumber\\
&=& \frac{i}{2}\left(\begin{array}{cccc}
                -\psi_{n\uparrow}({\bm x}){\bar\psi}_{m\uparrow}({\bm y}) & -\psi_{n\uparrow}({\bm x}){\bar\psi}_{m\downarrow}({\bm y})
                       & -\psi_{n\uparrow}({\bm x}) \psi_{m\downarrow}({\bm y}) & \ \ \psi_{n\uparrow}({\bm x}) \psi_{m\uparrow}({\bm y}) \\
               -\psi_{n\downarrow}({\bm x}){\bar\psi}_{m\uparrow}({\bm y}) & -\psi_{n\downarrow}({\bm x}){\bar\psi}_{m\downarrow}({\bm y})
                       & -\psi_{n\downarrow}({\bm x})\psi_{m\downarrow}({\bm y}) & \ \ \psi_{n\downarrow}({\bm x})\psi_{m\uparrow}({\bm y}) \\
                 \ \ {\bar\psi}_{n\downarrow}({\bm x}){\bar\psi}_{m\uparrow}({\bm y}) & \ \ {\bar\psi}_{n\downarrow}({\bm x}){\bar\psi}_{m\downarrow}({\bm y})
                      & \ \ {\bar\psi}_{n\downarrow}({\bm x})\psi_{m\downarrow}({\bm y}) & - {\bar\psi}_{n\downarrow}({\bm x})\psi_{m\uparrow}({\bm y})\\
                - {\bar\psi}_{n\uparrow}({\bm x}){\bar\psi}_{m\uparrow}({\bm y}) & -{\bar\psi}_{n\uparrow}({\bm x}){\bar\psi}_{m\downarrow}({\bm y})
                      & -{\bar\psi}_{n\uparrow}({\bm x})\psi_{m\downarrow}({\bm y}) & \ \ {\bar\psi}_{n\uparrow}({\bm x})\psi_{m\uparrow}({\bm y})
                  \end{array}\right)
\label{eq:2.8}
\eea
\end{widetext}
and its Fourier transform
\bse
\label{eqs:2.9}
\be
B_{nm}({\bm k},{\bm p}) = \frac{1}{V} \int d{\bm x}\,d{\bm y}\ e^{-i{\bm k}\cdot{\bm x} + i{\bm p}\cdot{\bm y}}\,
                                          B_{nm}({\bm x},{\bm y}).
\label{eq:2.9a}
\ee
The $4\times 4$ matrix $B_{nm}({\bm k},{\bm p})$ can be expanded in the spin-quaternion
basis defined above,
\be
B_{nm}({\bm k},{\bm p}) = \sum_{i,r=0}^3 {^i_r B}_{nm}({\bm k},{\bm p})\,(\tau_r\otimes s_i)\ .
\label{eq:2.9b}
\ee
It is further useful to define
\be
B_{nm}({\bm k};{\bm q}) = B_{nm}({\bm k}+{\bm q}/2,{\bm k}-{\bm q}/2)\ ,
\label{eq:2.9c}
\ee
\ese
with analogous definitions for other objects that depend on two wavevectors.
All bilinear products of the fermion fields ${\bar\psi}$ and $\psi$ can be written in terms of $B$,
and in particular all terms in the interacting part of the action, Eq.\ (\ref{eq:2.4b}), can be written
in terms of products of the $B$. An inspection shows that in the spin-quaternion basis,
Eq.\ (\ref{eqs:2.7}), the matrix elements ${_{r=0,3}^{\ \ \ \ \ i}B}$ and ${^{\ \ \ \ \ i}_{r=1,2}B}$ describe the 
particle-hole and particle-particle channels, respectively, and the ${^{i=0}_{\ \ \, r} B}$ describe the
spin-singlet channel while the ${^{i=1,2,3}_{\ \ \ \ \ \ \, r} B}$ describe the spin-triplet channel.
Taking moments of $B_{nm}({\bm k};{\bm q})$ with respect to ${\bm k}$ generates 
variables in different angular momentum channels. For instance,
\bse
\label{eq:2.10}
\be
B_{nm}^{(0)}({\bm q}) = \sum_{\bm k} B_{nm}({\bm k};{\bm q})
\label{eq:2.10a}
\ee
defines s-wave or density degrees of freedom,
\be
{\bm B}_{nm}^{(1)}({\bm q}) = \sum_{\bm k} {\bm k}\, B_{nm}({\bm k};{\bm q})
\label{eq:2.10b}
\ee
\ese
defines current degrees of freedom, etc.

We note that the introduction of bispinors, and of matrices with spin-quaternion valued 
matrix elements, is necessary in order to handle both particle-particle and particle-hole 
degrees of freedom within the same framework, as it allows to form bilinear products
${\bar\psi}{\bar\psi}$ and $\psi\psi$ in addition to ${\bar\psi}\psi$,
see Eq.\ (\ref{eq:2.8}). For purposes that involve the particle-hole
channel only one can restrict oneself to ordinary spinors, in which
case the matrix $B$, Eq.\ (\ref{eq:2.8}), becomes a $2\times 2$ matrix. Another observation is
that $B$ constitutes an overcomplete representation of bilinear fermion degrees of freedom since
all of the matrix elements of B are not independent. We will come back to this point in Sec.\ 
\ref{subsubsec:II.B.3} below.

\subsubsection{$Q$-matrix field theory}
\label{subsubsec:II.B.2}

Our next step is to constrain the matrices $B$ in the interaction terms to a classical matrix
field $Q$ by means of a Lagrange multiplier field $\tilde\Lambda$. The fermion fields then
enter the action only bilinearly and can be integrated out excatly. This way we obtain an effective
action ${\cal A}$ that depends on $Q$ and $\tilde\Lambda$ according to
\bea
Z &=& \int D[{\bar\psi},\psi]\ e^{S[{\bar\psi},\psi]} = \int D[\eta]\ e^{S[\eta]}
\nonumber\\
&=& \int D[\eta]\ e^{S[\eta]} \int D[Q,{\tilde\Lambda}]\ e^{\Tr[{\tilde\Lambda}(Q-B)]}
\nonumber\\
&=& \int D[Q,{\tilde\Lambda}]\ e^{{\cal A}[Q,{\tilde\Lambda}]}\ .
\label{eq:2.11}
\eea
Here and it what follows $\Tr$ denotes a trace over all degrees of freedom, including
the continuous position in real space, while by $\tr$ we will denote a  trace over all
discrete degrees of freedom that are not explicitly shown. 

In order to describe phenomena in angular-momentum channels higher than the
s-wave one, or even in order to keep non-s-wave interaction constants, one obviously
needs to apply this procedure to the phase-space variables defined in Eqs.\ (\ref{eq:2.8}, \ref{eqs:2.9}).
As we will see later, this leads to a local field theory that allows
for a systematic loop expansion about a saddle-point solution that describes a Fermi
liquid. For certain purposes that require a density-channel interaction only it is technically 
advantageous to formulate the theory in terms of density variables, Eq.\ (\ref{eq:2.10a}), 
even though this leads to a nonlocal theory, i.e., one where the vertices diverge in the
limit of small wave numbers. The starting point for such a formulation was given in Ref.\ 
\onlinecite{Belitz_Kirkpatrick_1997}, and we will revisit and further develop this
method in Sec.\ \ref{subsec:IV.E}.

Application of the procedure sketched in Eq.\ (\ref{eq:2.11}) to the phase space
variables, Eqs.\ (\ref{eq:2.8}, \ref{eqs:2.9}) yields the following formal expression for the 
effective action:
\bse
\label{eqs:2.12}
\be
{\cal A}[Q,{\tilde\Lambda}] = {\cal A}_0 + \Tr \left({\tilde\Lambda}^T Q\right)
   + {\cal A}_{\text{int}}[Q]\ .
\label{eq:2.12a}
\ee
Here
\be
{\cal A}_0 = \frac{1}{2}\,\Tr \ln G^{-1}\ ,
\label{eq:2.12b}
\ee
where
\be 
G^{-1} = G_0^{-1} - i{\tilde\Lambda}
\label{eq:2.12c}
\ee
is the inverse Green operator, with
\bea
\left(G_0^{-1}\right)_{nm}({\bm x},{\bm y}) &=& \delta_{nm}\,\left(\tau_0\otimes s_0\right)
   \,\delta({\bm x}-{\bm y})\,
\nonumber\\
&&\hskip -10pt \times \left[i\omega_n + {\bm\nabla}^2/2\me + \mu\right]
\label{eq:2.12d}
\eea
\ese
the bare Green operator. If a nontrivial band structure is desired, $-{\bm\nabla}^2/2\me$ should
be replaced by an appropriate energy function $\epsilon({\bm\nabla})$, see Eqs.\ (\ref{eqs:2.2}).
Rewriting the interaction part of the action, Eq.\ (\ref{eq:2.4b}), in
terms of the $B$, and constraining the latter to $Q$ by means of the functional delta-constraint
we find
\bse
\label{eqs:2.13}
\be
{\cal A}_{\text{int}} = {\cal A}_{\text{int}}^{(0,\text{s})} + {\cal A}_{\text{int}}^{(0,\text{t})}
     + {\cal A}_{\text{int}}^{(1,\text{t})}
\label{eq:2.13a}
\ee
where
\begin{widetext}
\bea
{\cal A}_{\text{int}}^{(0,\text{s})} &=& \frac{T\,\Gamma_{\text{s}}^{(0)}}{2V} \sum_{r,s=0,3} (-)^r
    \sum_{n_1,n_2\atop n_3,n_4} \delta_{n_1-n_2,n_4-n_3} \sum_{{\bm k},{\bm p}} 
         {\sum_{\bm q}}^{\prime}
   \tr\Bigl((\tau_r \otimes s_0)\,Q_{n_1n_2}({\bm k};{\bm q})\Bigr) \Bigl(\tau_r \otimes s_0)\, 
   Q_{n_3n_4}({\bm p};-{\bm q})\Bigr)
\nonumber\\
\label{eq:2.13b}\\
{\cal A}_{\text{int}}^{(0,\text{t})} &=& \frac{T\,\Gamma_{\text{t}}^{(0)}}{2V} \sum_{r,s=0,3} (-)^r
    \sum_{i=1}^3
    \sum_{n_1,n_2\atop n_3,n_4} \delta_{N_1-n_2,n_4-n_3} \sum_{{\bm k},{\bm p}} 
         {\sum_{\bm q}}^{\prime}
   \tr\Bigl((\tau_r \otimes s_i)\,Q_{n_1n_2}({\bm k};{\bm q})\Bigr) \Bigl(\tau_r \otimes s_i)\, 
   Q_{n_3n_4}({\bm p};-{\bm q})\Bigr)
\nonumber\\
\label{eq:2.13c}\\
{\cal A}_{\text{int}}^{(1,\text{t})} &=& \frac{T\,\Gamma_{\text{t}}^{(1)}}{2V} \sum_{r,s=0,3} (-)^r
    \sum_{i=1}^3
    \sum_{n_1,n_2\atop n_3,n_4} \delta_{n_1-n_2,n_4-n_3} \sum_{{\bm k},{\bm p}} 
       {\bm k}\cdot{\bm p}\  {\sum_{\bm q}}^{\prime}
   \tr\Bigl((\tau_r \otimes s_i)\,Q_{n_1n_2}({\bm k};{\bm q})\Bigr) \Bigl(\tau_r \otimes s_i)\, 
   Q_{n_3n_4}({\bm p};-{\bm q})\Bigr)
\nonumber\\
\label{eq:2.13d}
\eea
\end{widetext}
\ese

\subsubsection{Symmetry properties, and representation of observables}
\label{subsubsec:II.B.3}

We now derive some useful symmetry properties of the $Q$-matrices. $B$ as defined
in Eq.\ (\ref{eq:2.8}) is self-adjoint under the operation defined in Eq.\ (\ref{eq:2.6b}).
$Q$ inherits this property, so we have
\bse
\label{eqs:2.14}
\be
Q^+ = C^T\,Q^T\,C = Q\ .
\label{eq:2.14a}
\ee
In the spin-quaternion basis defined in Eq.\ (\ref{eqs:2.7}) this implies
\bea
{^i_r Q}_{nm}({\bm x},{\bm y}) &=& \fourchoose{+}{+}{+}{-}{r}\,\fourchoose{+}{-}{-}{-}{i}\,
                                                 {^i_r Q}_{mn}({\bm y},{\bm x})\ ,
\nonumber\\
\label{eq:2.14b}\\
{^i_r Q}_{nm}({\bm k},{\bm p}) &=&  \fourchoose{+}{+}{+}{-}{r}\,\fourchoose{+}{-}{-}{-}{i}\,
                                                 {^i_r Q}_{mn}(-{\bm p},-{\bm k})\ ,
\nonumber\\
\label{eq:2.14c}\\
{^i_r Q}_{nm}({\bm k};{\bm q}) &=& \fourchoose{+}{+}{+}{-}{r}\,\fourchoose{+}{-}{-}{-}{i}\,
                                                 {^i_r Q}_{mn}(-{\bm k};{\bm q})\ .
\nonumber\\
\label{eq:2.14d}
\eea
\ese
Here the symbols $\fourchoose{+}{+}{+}{-}{r}$ etc.
denote a factor of $+1$ for $r = 0$ and a factor of $-1$ for $r = 1,2,3$, and analogously 
for $i$. We have made use of these relations in order to write
the interacting part of the action in the form of Eqs.\ (\ref{eqs:2.13}). They imply
that all of the $Q$-matrix elements are not independent. In a model with $N$
Matsubara frequencies, only $N(N+1)/2$ matrix elements are independent. We
will later choose these to be the ones with $n \geq m$. 

In terms of the $Q$-matrices the observables listed in Eqs.\ (\ref{eqs:2.5}) are given by
\bse
\label{eqs:2.15}
\bea
n({\bm q},i\Omega_n) &=& \sum_m \sum_{r=0,3} (-)^{3r/2}\, 
   \sum_{\bm k} {^0_r Q}_{m,m+n}({\bm k};{\bm q})\ ,
\nonumber\\
\label{eq:2.15a}\\
j_{\alpha}({\bm q},i\Omega_n) &=& \sum_m \sum_{r=0,3} (-)^{3r/2}\, 
   \sum_{\bm k} k_{\alpha}\, {^0_r Q}_{m,m+n}({\bm k};{\bm q})\ ,
\nonumber\\
\label{eq:2.15b}
\eea
\bea
n_{\text{s}}^i({\bm q},i\Omega_n) &=& \sum_m \sum_{r=0,3} (-)^{3r/2}\,  
   \sum_{\bm k} {^0_r Q}_{m,m+n}({\bm k};{\bm q})\ ,
\nonumber\\
\label{eq:2.15c}
\eea
\bea
j_{\text{s}}^{i\alpha}({\bm q},i\Omega_n) &=& \sum_m \sum_{r=0,3} (-)^{3r/2}\, 
   \sum_{\bm k} k_{\alpha}\, {^0_r Q}_{m,m+n}({\bm k};{\bm q})\ .
\nonumber\\
\label{eq:2.15d}
\eea
\ese
These are just examples of spin-singlet and spin-triplet observables in the $\ell = 0$ and
$\ell = 1$ channels, respectively. Clearly, and desired observables can be expressed in terms
of the $Q$.

\subsubsection{Correlation functions}
\label{subsubsec:II.B.4}

Physical correlation functions can be written in terms of $Q$-correlation functions by keeping
appropriate source terms in the action while performing the transformation to the $Q$-matrix
variables. For instance, for the Green function $G_{n\sigma}({\bm x}-{\bm y}) = \langle\psi_{n\sigma}({\bm x})\,
{\bar\psi}_{n\sigma}({\bm y})\rangle$ we obtain
\be
G_{n\sigma}({\bm x}-{\bm y}) = \frac{i}{2}\,\tr\Bigl[\big((\tau_0+i\tau_3)\otimes(s_0\mp\sigma i s_3)\bigr)\langle
   Q_{nn}({\bm x},{\bm y})\rangle\Bigr]\ ,
\label{eq:2.16}
\ee
with $\sigma =\ \uparrow,\downarrow\ \equiv +1,-1$.
This can also be read off of Eq.\ (\ref{eq:2.8}) directly by keeping in mind the isomorphism between
$B$ and $Q$. For the density of states $N$ as a function of the distance $\omega$ from the
Fermi surface this implies
\be
N(\omega) = \frac{4}{\pi}\,\Re \langle{^0_0 Q}_{nn}({\bm x},{\bm x})\rangle\Bigr\vert_
{i\omega_n \to \omega + i0}\ ,
\label{eq:2.17}
\ee
where we have used the symmetry properties expressed in Eqs.\ (\ref{eqs:2.14}).
Similarly, the number density susceptibility $\chi$, and the spin density susceptibility 
tensor $\chi_{\text{s}}^{ij}$,
can be written
\bse
\label{eqs:2.18}
\bea
\chi({\bm q},i\Omega_n) &=& 16\,T\!\! \sum_{m_1,m_2} \sum_{r=0,3} \sum_{{\bm k},{\bm p}} \left\langle
   {^0_r(}\Delta Q)_{m_1-n,m_1}({\bm k};{\bm q})\right.
\nonumber\\
&&\hskip 27pt \left.\times {^i_r(}\Delta Q)_{m_2,m_2+n}({\bm p};-{\bm q})\right\rangle\ ,
\label{eq:2.18a}
\eea
\bea
\chi_{\text{s}}^{ij}({\bm q},i\Omega_n) &=& 16\,T\!\! \sum_{m_1,m_2} \sum_{r=0,3} \sum_{{\bm k},{\bm p}} \left\langle
   {^i_r(}\Delta Q)_{m_1-n,m_1}({\bm k};{\bm q})\right.
\nonumber\\
&&\hskip 27pt \left.\times {^j_r(}\Delta Q)_{m_2,m_2+n}({\bm p};-{\bm q})\right\rangle\ ,
\label{eq:2.18b}
\eea
\ese
Here $\Delta Q = Q - \langle Q \rangle$, and, in
Eq.\ (\ref{eq:2.18b}), $i,j = 1,2,3$. Other correlation functions can be expressed analogously.

\bigskip
\section{Identification of soft modes: Broken symmetry, and a Ward identity}
\label{sec:III}

We are interested in separating the degrees of freedom represented by the $Q$-matrices into
soft and massive modes. To identify the soft modes we derive a Ward identity that relates
two-point $Q$-correlation functions to other quantities. We first do so for noninteracting
electrons and then discuss the effects of interactions.

\subsection{A Ward identity for noninteracting electrons}
\label{subsec:III.A}

Let us consider transformations of the bispinors defined in Eqs.\ (\ref{eqs:2.6}),
\bse
\label{eqs:3.1}
\be
\eta_n({\bm x}) \to \int d{\bm y}\ {\hat T}^{(\pm)}_{nm}({\bm x},{\bm y})\,\eta_m({\bm y})\ ,
\label{eq:3.1a}
\ee
where the operator ${\hat T}^{(\pm)}$ defines non-local infinitesimal rotations in frequency
space,
\bea
{\hat T}^{(\pm)}_{nm}({\bm x},{\bm y}) &=& (\tau_0 \otimes s_0)\,t^{(\pm)}_{nm}({\bm x},{\bm y})\ ,
\label{eq:3.1b}\\
t^{(\pm)}_{nm}({\bm x},{\bm y}) &=& \delta_{nm}\,\delta({\bm x}-{\bm y}) + \left[\delta_{nn_1}\delta_{mn_2}
   - \delta_{nn_2}\delta_{mn_1}\right]\,
\nonumber\\
&& \hskip 20pt \times\varphi_{\pm}({\bm x},{\bm y}) + O(\varphi^2)\ .
\label{eq:3.1c}
\eea
Here the
\bea
\varphi_{\pm}({\bm x},{\bm y}) &=& \frac{1}{2}\,\left[\phi({\bm x},{\bm y}) \pm \phi({\bm y},{\bm x})\right]
\nonumber\\
                                                   &=& \pm \varphi_{\pm}({\bm y},{\bm x})
\label{eq:3.1d}
\eea
\ese
are even and odd combinations, respectively, of a non-local rotation angle $\phi$. The matrices
${\hat T}^{\pm}$ obey $({\hat T}^{(\pm)} )^T = ({\hat T}^{(\pm)} )^{-1}$, and 
$({\hat T}^{(\pm)} )^T\,C\,{\hat T}^{(\pm)}  = C$ with $C$ the charge conjugation matrix defined below
Eq.\ (\ref{eq:2.6b}). For fixed $n_1$ and $n_2$ they
form an SO$(2)$ subgroup of a much larger symplectic group (Sp($8N,\C$) for
a system with $2N$ frequency labels; $N$ positive ones, including zero, and $N$ negative ones), see
Ref.\ \onlinecite{Belitz_Kirkpatrick_1997}. The $Q$-matrices transform as
\be
Q({\bm x},{\bm y}) \to (T^{(\pm)}\,Q\,(T^{(\pm)})^T\ ,
\label{eq:3.2}
\ee
Explicitly we find
\bse
\label{eqs:3.3}
\be
Q_{nm}({\bm x},{\bm y}) \to Q_{nm}({\bm x},{\bm y}) + \delta Q_{nm}({\bm x},{\bm y})
\label{eq:3.3a}
\ee
with
\begin{widetext}
\be 
\delta Q_{nm}({\bm x},{\bm y}) = \int d{\bm z}\ \Bigl\{\varphi_{\pm}({\bm y},{\bm z})\,
   \left[\delta_{mn_1}\,Q_{nn_2}({\bm x},{\bm z}) \mp \delta_{mn_2}\,Q_{nn_1}({\bm x},{\bm z})\right]
   + \varphi_{\pm}({\bm x},{\bm z})\,\left[\delta_{nn_1}\,Q_{n_2m}({\bm z},{\bm y}) \mp \delta_{nn_2}\,
   Q_{n_1m}({\bm z},{\bm y})\right]\Bigr\}\ ,
\label{eq:3.3b}
\ee
and in particular 
\be
\delta Q_{n_1n_2}({\bm x},{\bm y}) = \int d{\bm z}\ \Bigl[\varphi_{\pm}({\bm x},{\bm z})\,Q_{n_2n_2}({\bm z},{\bm y}) 
  - Q_{n_1n_1}({\bm x},{\bm z})\,\varphi_{\pm}({\bm z},{\bm y})\Bigr]
\label{eq:3.3c}
\ee
\end{widetext}
\ese
The Lagrange multiplier field ${\tilde\Lambda}$ transforms as $Q$ does, on account of the bilinear
coupling between the two.  Of the three terms in the action, Eqs.\ (\ref{eqs:2.12}), the second one is
invariant under these transformations, but ${\cal A}_0$ and ${\cal A}_{\text{int}}$ are not. Focusing
on noninteracting electrons for the time being, we find ${\cal A}_0 \to {\cal A}_0 + \delta{\cal A}_0$
with
\bea
\delta{\cal A}_0 &=& \sum_{{\bm k},{\bm q}} \left[i\Omega_{n_1-n_2} + {\bm k}\cdot{\bm q}/\me \right]\, 
   \tr G_{n_2n_1}({\bm k};{\bm q})
\nonumber\\
&&\times\varphi_{\pm}({\bm k}-{\bm q}/2,{\bm k}+{\bm q}/2)\ .
\label{eq:3.4}
\eea
Here $G$ is the inverse of the Green operator defined in Eq.\ (\ref{eq:2.12c}).

A Ward identity can now be derived by standard techniques.\cite{Zinn-Justin_1996} We introduce a matrix 
source field $J$ for the $Q$, consider the generating functional
\be
Z[J] = \int D[Q,{\tilde\Lambda}]\ e^{{\cal A}_0 + \Tr ({\tilde\Lambda}^T Q) + \Tr(JQ)}\ ,
\label{eq:3.5}
\ee
perform the infinitesimal rotation defined by $\varphi_{\pm}$, differentiate with respect to $J$,
and put $J=0$. This way we obtain a Ward identity
\be
\left\langle \delta{\cal A}_0\,Q_{n_1n_2}({\bm x},{\bm y}) \right\rangle_{{\cal A}_0}
   + \left\langle \delta Q_{n_1n_2}({\bm x},{\bm y}) \right\rangle_{{\cal A}_0} = 0\ .
\label{eq:3.6}
\ee
From Eqs.\ (\ref{eq:3.4}) and (\ref{eq:3.6}) we see that this relates correlation functions
of the structure $\langle \tr G\,Q\rangle$ to $\langle Q\rangle$. The former can be rewritten
in terms of $\langle Q\,Q\rangle$ by generalizing the generating functional given in Eq.\ (\ref{eq:3.5}).
Since the $Q$ are isomorphic to $B$, Eq.\ (\ref{eq:2.8}), we can write the source term
$JQ = xJQ + (1-x)JB$ with an arbitrary real number $x$. The generating functional then becomes
\begin{widetext}
\be
Z[J] = \int D[Q,{\tilde\Lambda}]\ e^{\frac{1}{2}\,\Tr\ln [G^{-1} + i(1-x)J^T] + x\,\Tr(JQ) + \Tr({\tilde\Lambda}^T Q)}\ .
\label{eq:3.7}
\ee
Note that this is independent of $x$, and that by choosing $x=1$ we recover Eq.\ (\ref{eq:3.5}). By
differentiating with respect to $J$, choosing $x=0$ and $x=1$, respectively, and putting $J=0$
we obtain an identity
\bse
\label{eqs:3.8}
\be
\langle G_{n_2n_1}({\bm x}_2,{\bm x}_1)\rangle = -2i\,\langle Q_{n_1n_2}({\bm x}_1,{\bm x}_2)\rangle\ .
\label{eq:3.8a}
\ee
Differentiating twice with respect to $J$ we find
\be
\langle G_{n_2n_1}({\bm x}_2,{\bm x}_1)\,Q_{n_3n_4}({\bm x}_3,{\bm x}_4)\rangle = -2i\,\langle 
   Q_{n_1n_2}({\bm x}_1,{\bm x}_2)\,Q_{n_3n_4}({\bm x}_3,{\bm x}_4) \rangle\ .
\label{eq:3.8b}
\ee
\ese
We now differentiate Eq.\ (\ref{eq:3.6}) with respect to $\phi$ in Eq.\ (\ref{eq:3.1d}). This yields two
identities, one for $\varphi_+$ and one for $\varphi_-$. Adding them yields the Ward identity in
its final form:
\be
D_{n_1n_2,n_3n_4}({\bm k},{\bm p};{\bm q}) \equiv
\left\langle {^0_0 Q}_{n_1n_2}({\bm k};{\bm q})\,{^0_0 Q}_{n_3n_4}({\bm p};{\bm q}) \right\rangle = \frac{i}{8}\,
   \delta_{{\bm k},{\bm p}}\ \delta_{n_1n_3}\,\delta_{n_2n_4}\,
   \frac{\left\langle{^0_0 Q}_{n_1n_1}({\bm k}-{\bm q}/2)\right\rangle - \left\langle{^0_0 Q}_{n_2n_2}({\bm k}+{\bm q}/2)\right\rangle}
   {i\Omega_{n_1-n_2} + {\bm k}\cdot{\bm q}/\me}\ .
\label{eq:3.9}
\ee

\subsection{Soft modes in noninteracting electron systems}
\label{subsec:III.B}

Let us discuss the Ward identity for noninteracting electrons, Eq.\ (\ref{eq:3.9}).  
By using Eq.\ (\ref{eq:3.8a}) we see that the right-hand side of Eq.\ (\ref{eq:3.9}) is determined by
\be
-2i\,\langle Q_{nn}({\bm k}\pm{\bm q}/2)\rangle = \langle G_{nn}({\bm k} + {\bm q}/2)\rangle = \frac{1}
   {i\omega_n - \xi_{{\bm k}\pm{\bm q}/2}}
\to \mp i\pi\,\delta(\xi_{{\bm k}\pm{\bm q}/2}) - 1/\xi_{{\bm k}\pm{\bm q}/2}\ .
\label{eq:3.10}
\ee
\end{widetext}
 Here $\xi_{\bm k} = {\bm k}^2/2\me - \mu$, and the last term on indicates the limiting value of the
Green function as $\omega_n$ approaches zero from above or below, respectively. We see that if
$n_1$ and $n_2$ have opposite signs, then the numerator on the right-hand side of Eq.\ (\ref{eq:3.9})
goes to zero as $\Omega_{1-2} \to 0$ and ${\bm q} \to 0$, as does the denominator, whereas the
numerator remains finite if $n_1$ and $n_2$ have opposite signs. In the latter case, the leading
term for small $\Omega_{1-2}$ and small ${\bm q}$ is
\bea
\left\langle {^0_0 Q}_{n_1n_2}({\bm k};{\bm q})\,{^0_0 Q}_{n_1n_2}({\bm p};-{\bm q}) \right\rangle &=& 
\nonumber\\
&& \hskip-140pt  \delta_{{\bm k},{\bm p}}\,\frac{1}{16}\,
   \frac{2\pi i\,\delta(\xi_{\bm k})\,\sgn \Omega_{n_1-n_2}}{i\Omega_{n_1-n_2} + {\bm k}\cdot{\bm q}/\me}
   +\ O(1)\ .
\label{eq:3.11}
\eea
We see that there is an infinite number of soft modes that can be obtained by taking all possible
moments of Eq.\ (\ref{eq:3.11}) with respect to the center-of-mass wave vector ${\bm k}$. This
identifies the matrix elements $Q_{nm}$ with $nm<0$ as soft degrees of freedom, whereas the
$Q_{nm}$ with $nm>0$ are massive. Physically, the soft modes are particle-hole excitations with
a linear frequency-momentum relation. In the absence of interactions they are massless at
nonzero temperature as well as at $T=0$.

Equation (\ref{eq:3.11}) is a generalization of the Ward identity derived in Ref.\ \onlinecite{Belitz_Kirkpatrick_1997},
which corresponded to the zeroth moment at ${\bm q}=0$,
\be
\left\langle{^0_0 Q}_{n_1n_2}^{(0)}({\bm q})\,{^0_0 Q}_{n_1n_2}^{(0)}(-{\bm q})\right\rangle\Bigr\vert_{{\bm q}=0}
   = \frac{\pi \NF}{8\vert\Omega_{n_1-n_2}\vert}\ .
\label{eq:3.12}
\ee
Here $Q^{(0)}({\bm q}) = \sum_{\bm k} Q({\bm k};{\bm q})$ is defined in analogy to $B^{(0)}$, Eq.\ (\ref{eq:2.10a}), 
and $\NF$ is the free-electron
density of states per spin at the Fermi level. Equation (\ref{eq:3.12}) is the clean analog of an identity 
that has been discussed before for disordered electrons.\cite{Maleev_Toperverg_1975,
Vollhardt_Woelfle_1980, Schaefer_Wegner_1980, McKane_Stone_1981, Pruisken_Schaefer_1982}
In the presence of quenched disorder this is the {\em only}
soft mode; all higher moments acquire a mass, see Appendix \ref{app:A}. This is the reason why it
is much harder to construct a soft-mode theory for clean electrons than for disordered ones: In the
former case there are many more soft modes.

The Ward identity shows only that the matrix elements ${^0_0 Q}_{nm}$ with $nm < 0$ are soft.
However, it is easy to see that the same statements holds for the ${^i_r Q}$ with arbitrary values of
$i$ and $r$. The reason is that the $\langle {^i_r Q}\,{^i_r Q}\rangle$ correlation functions are
related to $\langle {^0_0 Q}\,{^0_0 Q}\rangle$ by means of symmetries that are not broken.
This argument is identical to the one give in Ref.\ \onlinecite{Belitz_Kirkpatrick_1997} and we
do not repeat it here.

\subsection{Interaction effects}
\label{subsec:III.C}

We now need to address the question of what happens if the electron-electron interactions are
taken into account. To this end we need to add the variation of the interacting part of the
action, $\delta{\cal A}_{\text{int}}$, under the transformation given in Eqs.\ (\ref{eqs:3.1}),
to Eq.\ (\ref{eq:3.6}). $\delta{\cal A}_{\text{int}}$ is quadratic in $Q$, and the Ward identity
now relates the 2-point function on the left-hand side of Eq.\ (\ref{eq:3.9}) to a 1-point
function and a 3-point function. The technical details are cumbersome and have been
derived in Ref.\ \onlinecite{Belitz_Kirkpatrick_1997}, but to show that the particle-hole
excitations identified in Eq.\ (\ref{eq:3.9}) remain soft it suffices to discuss the structure
of the generalized Ward identity, as was shown in Ref.\ \onlinecite{Kirkpatrick_Belitz_2002}.
In what follows we recapitulate and adapt the argument give there.
Equation (\ref{eq:3.9}) gets generalized to
\bse
\label{eqs:3.13}
\bea
&-&\hskip-5pt 8i\,(i\Omega_{n_1-n_2} + {\bm k}\cdot{\bm q}/\me^*)\,D_{n_1n_2,n_3n_4}({\bm k},{\bm p};{\bm q})\qquad\quad
\nonumber\\
&& \hskip -5pt = \delta_{n_1n_3}\,\delta_{n_2n_4}\,\delta_{{\bm k},{\bm p}}\,N_{n_1n_2}({\bm k},{\bm q}) -
    W_{n_1n_2,n_3n_4}({\bm k},{\bm p};{\bm q})\ .
\nonumber\\
\label{eq:3.13a}
\eea
Here $\me^*$ is the effective electron mass that is renormalized by the Fermi-liquid parameter
$F_{\text{s}}^1$, and
\be
N_{n_1n_2}({\bm k},{\bm q}) = \pi\,Z_{\bm k}\,\delta(\xi_{\bm k})\,\sgn \Omega_{n_1-n_2}
\label{eq:3.13b}
\ee
\ese
with $Z_{\bm k}$ the quasi-particle weight that reflects the change from free-electron
Green functions to physical ones. $W$ is the contribution from $\delta{\cal A}_{\text{int}}$. General
considerations show that there are two contributions to $D$ that are characterized by
different frequency structures,
\bea
D_{n_1n_2,n_3n_4}({\bm k},{\bm p};{\bm q}) &=& \delta_{n_1n_3}\,\delta_{n_2n_4}\,
     D_{n_1n_2}^{(\text{dc})}({\bm k},{\bm p};{\bm q})
\nonumber\\
   && \hskip -20pt+ \delta_{n_1-n_2,n_3-n_4}\,D_{n_1n_2,n_3n_4}^{(\text{c})}({\bm k},{\bm p};{\bm q})\ ,
\nonumber\\
\label{eq:3.14}
\eea
and analogously for $W$. Here the superscripts $(dc)$ and $(c)$, refer to the
disconnected and connected contributions, respectively, in a representation in terms of
fermionic diagrams. These contributions must respect the Ward identity separately, and
either one diverging in the limit of small frequencies and waves numbers will ensure
the existence of soft modes. It therefore suffices to discuss $D^{(\text{dc})}$. The Ward
identity for the latter reads
\bea
-8i\,(i\Omega_{n_1-n_2} + {\bm k}\cdot{\bm q}/\me^*)\,D^{(\text{dc})}_{n_1n_2}({\bm k},{\bm p};{\bm q}) &=& 
\nonumber\\&&\hskip -170pt
     \delta_{{\bm k},{\bm p}}\,N_{n_1n_2}({\bm k},{\bm q})  - W^{(\text{dc})}_{n_1n_2}({\bm k},{\bm p};{\bm q})\ .
\label{eq:3.15}
\eea
Now analytically continue to real frequencies $\Omega \in \R$, $i\Omega_{n_1-n_2} \to \Omega + i0$, and consider
the limit $\Omega \to 0$, $q \to 0$. The only way for $D^{(\text{dc})}$ at $t=0$ to remain finite in this 
limit is for $N$ and $W^{(\text{dc})}$ at $\Omega = q = 0$ to cancel. However this is in general
not possible: $N$ approaches a nonzero limit for $\Omega \to 0$, $q \to 0$, and vanishing
interactions, whereas $W$ vanishes for noninteracting electrons. These two objects are thus
different functions of the interaction, and their values at zero frequency and wave number can
vanish at most for special values of the interaction. Notice that no such a cancellation is necessary
to give $D^{(\text{dc})}$ a mass at nonzero temperature. For simplicity, consider the zeroth moment
with respect to ${\bm k}$ and ${\bm p}$ only, and put ${\bm q}=0$. Then we have
\be
-8i\Omega D^{(\text{dc})} = N - W^{(\text{dc})}\ .
\label{eq:3.16}
\ee
Now let 
\be
D^{(\text{dc})} \propto \frac{1}{\Omega + i/\tau_{\phi}}\ ,
\label{eq:3.17}
\ee
with $1/\tau_{\phi}$ a phase relaxation rate. As long as $\Omega\tau_{\phi} \gg 1$ for $T \to 0$
and $\Omega \to 0$, the right-hand side of Eq.\ (\ref{eq:3.16}) can be expanded in powers
of $1/\Omega\tau_{\phi}$, with the leading contribution being a constant. As long as
$1/\tau_{\phi} \to 0$ for $T \to 0$ such a regime always exists.

We conclude that the soft modes identified in Sec.\ \ref{subsec:III.A}, namely, the
$Q_{nm}$ with $nm<0$, remain soft in the presence of interactions. The Ward identity
shows this for the matrix elements ${^0_0 Q}$ in the spin-quaternion basis, but 
additional symmetries of the action ensure that it is true for all of the 
${^i_r Q}$.\cite{Belitz_Kirkpatrick_1997} This is consistent
with both perturbation theory and with Fermi-liquid theory, which ensures that the number 
and nature of the soft modes cannot change as long as the system remains in a Fermi-liquid 
phase. The symmetry is broken, and the Goldstone modes exist, as long as there is a
non-vanishing quasi-particle weight $Z_{\bm k}$. The only way the soft-mode spectrum 
can change is for the quasi-particle weight to vanish, which is to say that the system enters 
a non-Fermi-liquid state. We note that these considerations do not preclude using
the soft-mode theory to study a breakdown of the Fermi liquid, which is signalized by a
vanishing quasi-particle weight. They just indicate that a non-Fermi-liquid phase, where the
symmetry is restored, has different properties and requires a different effective theory than
the Fermi-liquid phase, but the theory we are about to construct will be valid everywhere in
the latter, up to any possible transition where the quasi-particle weight vanishes. This is
important for certain applications of the theory.\cite{Kirkpatrick_Belitz_2011b}

We finally note that while the right-hand side of Eq.\ (\ref{eq:3.11}) resembles the structure of the
Lindhard function, the soft modes are not related to particle-number conservation, and are
not restricted to the density channel. Indeed, the density susceptibility cannot even be constructed
from the left-hand side, as doing so would require {\em three} independent frequencies, rather than
two, as can be seen from Eq.\ (\ref{eq:2.18a}). Rather, the infinitely many soft modes are the
Goldstone modes of a spontaneously broken continuous symmetry, namely, the rotations in
frequency space expressed by Eqs.\ (\ref{eqs:3.1}).

\section{Effective soft-mode theory}
\label{sec:IV}

\subsection{Saddle-point solutions}
\label{subsec:IV.A}

We now return to the action, Eqs.\ (\ref{eqs:2.12}, \ref{eqs:2.13}), and consider saddle-point
solutions. There are many saddle-point solutions that have different symmetry properties
and describe different physical states. Here we give just two examples. The first one is a
saddle point that describes a Fermi liquid, and the second one reproduces the Stoner
theory of ferromagnetism. Saddle points that describe superconducting states, 
magnetic states with non-s-wave order parameters, or, if a nontrivial band structure is
taken into account, antiferromagnetic states, can be constructed analogously.

\subsubsection{Fermi-liquid saddle point}
\label{subsubsec:IV.A.1}

The saddle-point equations are obtained by minimizing the action with respect to
${\tilde\Lambda}$ and $Q$. They read
\bse
\label{eqs:4.1}
\bea
0 &=& -\,\frac{i}{2}\,\left(G_{mn}^{\text{sp}}({\bm y},{\bm x})\right)^T 
            + Q_{nm}^{\text{sp}}({\bm x},{\bm y})\ .
\label{eq:4.1a}\\
0 &=& {\tilde\Lambda}_{nm}^{\text{sp}}({\bm x},{\bm y}) 
          + \frac{\delta}{\delta Q_{nm}({\bm x},{\bm y})}\Bigr\vert_{Q^{\text{sp}}}\,
           {\cal A}_{\text{int}}[Q]\ .
\label{eq:4.1b}
\eea
\ese
To find a saddle-point solution that describes a Fermi liquid we make an ansatz
\bea
Q_{nm}^{\text{sp}}({\bm x},{\bm y}) &=& (\tau_0\otimes s_0)\,\delta_{nm}\,Q_{n}({\bm x}-{\bm y})\ ,
\nonumber\\
{\tilde\Lambda}_{nm}^{\text{sp}}({\bm x},{\bm y}) &=& (\tau_0\otimes s_0)\,\delta_{nm}\,
   \Lambda_{n}({\bm x}-{\bm y})\ .
\label{eq:4.2}
\eea
Performing the derivative in Eq.\ (\ref{eq:4.1a}) and  going into Fourier space we find 
$Q_n$ and $\Lambda_n$ as the solution of the equations
\bse
\label{eqs:4.3}
\bea
Q_n({\bm k}) &=& \frac{i}{2}\,\frac{1}{i\omega_n - \xi_{\bm k} - i\Lambda}\ ,
\label{eq:4.3a}\\
\Lambda_n({\bm k}) &\equiv& \Lambda = -4\Gamma_{\text{s}}^{(0)}\,\frac{1}{V}
   \sum_{\bm p} T \sum_m e^{i\omega_m 0}\,Q_m({\bm p})\ .
\nonumber\\
\label{eq:4.3b}
\eea
\ese
We see that $-2iQ_n({\bm k})$ equals the saddle-point Green function, in agreement
with Eq.\ (\ref{eq:3.8a}), with $\Lambda$ the self energy. The latter represents the 
spin-singlet interaction $\Gamma_{\text{s}}^{(0)}$ in Hartree-Fock approximation. 
The factor $e^{i\omega_m 0}$ in Eq.\ (\ref{eq:4.3b}) is the usual convergence factor
that resolves the ambiguity inherent in the equal-time Green function.\cite{Fetter_Walecka_1971}
This Hartree-Fock saddle-point solution is a generalization of the saddle-point solution
considered in Ref.\ \onlinecite{Belitz_Kirkpatrick_1997} to a non-local  Green function.

\subsubsection{Stoner saddle point}
\label{subsubsec:IV.A.2}

To illustrate the existence of other saddle points, and the general versatility of the theory,
we now consider a saddle-point ansatz
\bse
\label{eqs:4.4}
\bea
Q_{nm}^{\text{sp}}({\bm x},{\bm y}) &=& \delta_{12}\,\Bigl[(\tau_0\otimes s_0)\,G_{n}({\bm x}-{\bm y})
\nonumber\\
&&\hskip 20pt   + (\tau_3\otimes s_3)\,F_{n}({\bm x}-{\bm y})\Bigr]\ ,
\label{eq:4.4a}\\
{\tilde\Lambda}_{nm}^{\text{sp}}({\bm x},{\bm y}) &=& \delta_{nm}\,\Bigl[-(\tau_0\otimes s_0)\,i\Sigma_{n}({\bm x}-{\bm y})
\nonumber\\
&&\hskip 20pt   + (\tau_3\otimes s_3)\,i\Delta_{n}({\bm x}-{\bm y})\Bigr]\, .
\label{eq:4.4b}
\eea
\ese
The saddle-point equations (\ref{eqs:4.1}) now yield
\bse
\label{eqs:4.5}
\bea
G_n({\bm k}) &=& \frac{i}{4}\,\left[{\cal G}_n^+({\bm k}) + {\cal G}_n^-({\bm k})\right]\ ,
\label{eq:4.5a}\\
F_n({\bm k}) &=& \frac{i}{4}\,\left[{\cal G}_n^+({\bm k}) - {\cal G}_n^-({\bm k})\right]\ ,
\label{eq;4.5b}\\
\Sigma_n({\bm k}) &\equiv& \Sigma = -4i\Gamma_{\text{s}}^{(0)}\,\frac{1}{V}\sum_{\bm p}
   T\sum_m e^{i\omega_m0}\,G_m({\bm p})\ ,
\nonumber\\
\label{eq:4.5c}\\
\Delta_n({\bm k}) &\equiv& \Delta = -4i\Gamma_{\text{t}}^{(0)}\,\frac{1}{V}\sum_{\bm p}
   T\sum_m e^{i\omega_m0}\,F_m({\bm p})\ .
\nonumber\\
\label{eq:4.5d}
\eea
\ese
Here
\be
{\cal G}_n^{\pm}({\bm k}) = \frac{1}{i\omega_n - \xi_{\bm k} \pm \Delta - \Sigma}
\label{eq:4.6}
\ee
are Green functions whose self energy contributions describe the particle-hole spin-triplet
and spin-singlet interactions in Hartree-Fock approximation. This saddle-point solution is
the clean limit of the on considered in Ref.\ \onlinecite{Kirkpatrick_Belitz_2000}, generalized
to the case of non-local Green functions. It is equivalent to Stoner theory, which can be seen
as follows. $G_n({\bm k})$ and $F_n){\bm k})$ obey the equations
\bse
\label{eqs:4.7}
\bea
(i\omega_n - \xi_{\bm k} - \Sigma)\,{\cal G}_n({\bm k}) + \Delta\,{\cal F}_n({\bm k}) = 1\ ,
\label{eq:4.7a}\\
(i\omega_n - \xi_{\bm k} - \Sigma)\,{\cal F}_n({\bm k}) + \Delta\,{\cal G}_n({\bm k}) = 0\ .
\label{eq:4.7b}
\eea
\ese
If we absorb the constant spin-singlet self energy $\Sigma$ into a redefinition of the
chemical potential we obtain from Eqs.\ (\ref{eqs:4.7}) the equation of state in a form
that is familiar from Stoner theory,\cite{Moriya_1985}
\be
1 = -2\Gamma_{\text{t}}^{(0)}\,T\sum_n \frac{1}{V}\sum_{\bm k} \frac{1}{(i\omega_n - \xi_{\bm k})^2 - \Delta^2}\ ,
\label{eq:4.8}
\ee
which allows for a nonzero magnetization (which is proportional to $\Delta$) provided $\NF\Gamma_{\text{t}}^{(0)} > 1$.

\subsection{Expansion about the saddle point}
\label{subsec:IV.B}

In the remainder of this paper we consider fluctuations about the Fermi-liquid saddle
point derived in Sec. \ref{subsubsec:IV.A.1}; an application of the theory to
magnetically ordered states will be discussed elsewhere.\cite{us_tbp} 

\subsubsection{Gaussian fluctuations}
\label{subsubsec:IV.B.1}

We first consider Gaussian fluctuations about the Fermi-liquid saddle point. If we write
$Q = Q^{\text{sp}} + \delta Q$ and ${\tilde\Lambda} = {\tilde\Lambda}^{\text{sp}} +
 \delta{\tilde\Lambda}$, an expansion of the action, Eqs.\ (\ref{eqs:2.12}, \ref{eqs:2.13}),
yields
\be
{\cal A}[Q,{\tilde\Lambda}] = {\cal A}^{\text{sp}} + {\cal A}^{(2)} + \Delta{\cal A}\ ,
\label{eq:4.9}
\ee
where ${\cal A}^{\text{sp}}$ is the saddle-point action, ${\cal A}^{(2)}$ denotes the
Gaussian fluctuations, and $\Delta{\cal A}$ denotes the contributions that are of
cubic or higher order in the fluctuations. For the Gaussian part we find
\be
{\cal A}^{(2)} = \frac{1}{4}\,\Tr (G_{\text{sp}}\,\delta{\tilde\Lambda}\,G_{\text{sp}}
                        \,\delta{\tilde\Lambda}) + \Tr (\delta{\tilde\Lambda}^T \delta Q)
                        + {\cal A}_{\text{int}}[\delta Q]\ .
\label{eq:4.10}
\ee
Here
\bse
\label{eqs:4.11}
\be
\left(G_{\text{sp}}\right)_{nm}({\bm k},{\bm p}) = (\tau_0\otimes s_0)\,\delta_{nm}\,
   \delta_{{\bm k},{\bm p}}\,G_n({\bm k})
\label{eq:4.11a}
\ee
with
\be
G_n({\bm k}) = \frac{1}{i\omega_n - \xi_{\bm k} - i\Lambda}\ ,
\label{eq:4.11b}
\ee
\ese
where $\Lambda$ is the solution of Eqs.\ (\ref{eqs:4.3}), is the Green function in 
saddle-point approximation. With our local interaction amplitude $\Lambda$ amounts only
to a constant shift of the chemical potential, so $G_n({\bm k})$ can be taken
to be the free-electron Green function. 

We can decouple $\delta{\tilde\Lambda}$ and $\delta Q$ at the Gaussian level by
defining a field $\delta{\bar\Lambda}$ by
\bse
\label{eqs:4.12}
\be
{^i_r(}\delta{\bar\Lambda})_{12} = \frac{1}{2}\,\phi_{12}\ {^i_r(}\delta{\tilde\Lambda})_{12} 
       + {^i_r(}\delta{\bar Q})_{12}\ .
\label{eq:4.12a}
\ee
Here we have defined $(\delta{\bar\Lambda})_{12} \equiv (\delta{\bar\Lambda})_{n_1n_2}({\bm p}_1,{\bm p}_2)$,
and analogously for other fields, as well as
\be
{^i_r(}\delta{\bar Q})_{12} = {^i_r(}\delta Q)_{12}\,\fourchoose{+}{+}{-}{-}{r}\,\fourchoose{+}{-}{+}{-}{i}
\label{eq:4.12b}
\ee
and
\bea
\phi_{12} &=& G_1\,G_2 \equiv G_{n_1}({\bm p}_1)\,G_{n_2}({\bm p}_2)
\nonumber\\
&\equiv& \begin{cases} \Phi_{12} & \text{if $(n_1 + 1/2)(n_2 + 1/2) > 0$}\\
                              \varphi_{12} & \text{if $(n_1 + 1/2)(n_2 + 1/2) < 0$}\ .
        \end{cases}
\nonumber\\
\label{eq:4.12c}
\eea
The frequency-restricted objects $\Phi_{12}$ and $\varphi_{12}$ will be useful later. 
We also define 
\be
(\delta Q)_{12}^{\ddagger} = (\delta Q)_{n_1n_2}(-{\bm p}_1,-{\bm p}_2)\ , 
\label{eq:4.12d}
\ee
\ese
and
analogously for $\delta\Lambda$.
In terms of $\bar{\delta\Lambda}$ and $\delta Q$ the Gaussian action then takes the form
\bea
{\cal A}^{(2)} &=& -4\sum_{r,i} \fourchoose{+}{-}{-}{+}{r} \sum_{12} {^i_r(}\delta Q)_{12}\,
     \frac{1}{\phi_{12}}\,{^i_r(}\delta Q)_{12}^{\ddagger}
\nonumber\\
&&     + 4\sum_{r,i} \fourchoose{+}{-}{-}{+}{r} \sum_{12} {^i_r(}\delta{\bar\Lambda})_{12}\,
     \frac{1}{\phi_{12}}\,{^i_r(}\delta{\bar\Lambda})_{12}^{\ddagger}\ .
\nonumber\\
&& + {\cal A}_{\text{int}}[\delta Q]\ .
\label{eq:4.13}
\eea
Note that the Gaussian $\delta{\bar\Lambda}$ propagator equals minus the $\delta Q$
propagator\cite{contour_footnote} for noninteracting electrons,
\be
\langle \delta{\bar\Lambda}_{12}\,\delta{\bar\Lambda}_{34}\rangle_{{\cal A}^{(2)}} 
   = - \langle \delta Q_{12}\,\delta Q_{34} \rangle_{{\cal A}_0^{(2)}}\ ,
\label{eq:4.14}
\ee
with ${\cal A}^{(2)}_0$ the first two terms in Eq.\ (\ref{eq:4.10}) or (\ref{eq:4.13}). This
will be important for the structure of the effective soft-mode theory and the loop expansion.

\subsubsection{The fluctuation action in terms of independent variables}
\label{subsubsec:IV.B.2}

To go beyond Gaussian fluctuations we expand the $\Tr\ln$ term in the action, Eq.\ (\ref{eqs:2.12}),
in powers of $\delta{\tilde\Lambda}$ and express the result in terms of $\delta{\bar\Lambda}$ and
$\delta{\bar Q}$ by means of Eq.\ (\ref{eq:4.12a}). We find
\begin{widetext}
\bea
\Delta{\cal A} &=& -\sum_{m=3}^{\infty} \frac{i^m}{2m}\ \Tr (G_{\text{sp}}\,\delta{\tilde\Lambda})^m 
   \equiv \sum_{m=3}^{\infty} \Delta{\cal A}^{(m)}
\nonumber\\
&=& \frac{4i}{3} \sum_{1,2,3} G_1^{-1}\,G_2^{-1}\,G_3^{-1}\,\tr\left[(\delta{\bar\Lambda}_{12} - \delta{\bar Q}_{12})
   (\delta{\bar\Lambda}_{23} - \delta{\bar Q}_{23}) (\delta{\bar\Lambda}_{31} - \delta{\bar Q}_{31}) \right]
\nonumber\\
&& - 2 \sum_{1,2,3,4} G_1^{-1}\,G_2^{-1}\,G_3^{-1}\,G_4^{-1}\,\tr\left[(\delta{\bar\Lambda}_{12} - \delta{\bar Q}_{12})
   (\delta{\bar\Lambda}_{23} - \delta{\bar Q}_{23}) (\delta{\bar\Lambda}_{34} - \delta{\bar Q}_{34}) 
   (\delta{\bar\Lambda}_{41} - \delta{\bar Q}_{41})\right] + \ldots
\label{eq:4.15}
\eea
Notice that the vertices are given in terms of products of inverse Green functions, which are
well behaved in the limit of small frequencies of wave numbers. This is thus a local field theory,
in contrast to the formulation in terms of density modes in Ref.\ \onlinecite{Belitz_Kirkpatrick_1997}.

We now recall two features of the theory. First, all of the matrix elements of $Q$ and $\Lambda$ are not
independent. We can choose the $Q_{nm}$ with $n \geq m$ as the independent ones, and express those
with $n < m$ in terms of them by means of Eqs.\ (\ref{eqs:2.14}). Second, we recall from Sec.\ \ref{sec:III} 
that the $Q_{nm}$ are massive
degrees of freedom if $n$ and $m$ have the same sign, and soft ones if $n$ and $m$ have opposite
signs. We express these facts in our notation by choosing as independent matrix elements
\bse
\label{eqs:4.16}
\bea
(\delta{\bar Q})_{12} &=& \begin{cases} {\bar P}_{12} & \text{if $ n_1 \geq n_2$ $\wedge$ $(n_1 + 1/2)(n_2 + 1/2) > 0$}\\
                                                              q_{12} & \text{if $n_1 \geq 0 \wedge n_2 < 0$}\ ,
                                        \end{cases}
\nonumber\\
\label{eq:4.16a}\\
(\delta{\bar\Lambda})_{12} &=& \begin{cases} \Lambda_{12} & \text{if $n_1 \geq n_2$ $\wedge$ $(n_1 + 1/2)(n_2 + 1/2) > 0$)}\\
                                                                         \lambda_{12} & \text{if $n_1 \geq 0 \wedge n_2 < 0$}\ .
                                                   \end{cases}
\nonumber\\
\label{eq:4.16b}
\eea
\ese
In what follows we will absorb the frequency restrictions into the fields, so writing, e.g., 
$q_{12}$ implies that $n_1 \geq 0$ and $n_2 < 0$, etc.
For the Gaussian action we obtain
\bea
{\cal A}^{(2)} &=&  - 8\sum_{r,i} \fourchoose{+}{-}{-}{+}{r} \sum_{12} \frac{1}{\phi_{12}}\,
   \biggl[{^i_r q}_{12}\,{^i_r q}_{12}^{\ddagger} + I_{12}\, 
         {^i_r {\bar P}}_{12}\,{^i_r {\bar P}}_{12}^{\ddagger}
   - {^i_r \lambda}_{12}\,{^i_r \lambda}_{12}^{\ddagger}
   - I_{12}\,{^i_r \Lambda}_{12}\,{^i_r \Lambda}_{12}^{\ddagger}\biggr]
\nonumber\\
&&+ \frac{16T\Gamma_{\text{s}}^{(0)}}{V} \sum_{r=0,3} \sum_{1,2\atop 3,4} 
              \delta_{1-2,3-4} \biggl[{^0_r q}_{12}\,{^0_r q}_{34}^{\ddagger}
  + I_{12}\,{^0_r {\bar P}}_{12}\,{^0_r {\bar P}}_{34}^{\ddagger} 
   + {^0_r q}_{12}\,{^0_r {\bar P}}_{34}^{\ddagger} 
   + {^0_r {\bar P}}_{12}\,{^0_r q}_{34}^{\ddagger}\biggr]\ ,
\label{eq:4.17}
\eea
where $I_{12} = 1 - \delta_{n_1n_2}/2$.
Here we keep only the particle-hole spin-singlet interaction; the other interaction channels provide
analogous contributions to the Gaussian action. Notice that the massless and massive degrees of
freedom are decoupled in the noninteracting part of the action, but are coupled by the interaction 
via terms that are bilinear in $q$ and $\bar P$. 

The contributions to $\Delta{\cal A}$ can also be expressed in terms of ${\bar P}$, $q$, $\Lambda$, 
and $\lambda$. Simple combinatorics yield for the cubic term
\bea
\Delta{\cal A}^{(3)} &=& -4i\biggl[\Tr\left[G^{-1}\,({\bar P}-\Lambda)\,G^{-1}\,
   ({\bar P} + {\bar P}^+ -\Lambda - \Lambda^+)\,G^{-1}\,({\bar P}-\Lambda)^+\right]
\nonumber\\
&&\hskip 15pt +\Tr\left[G^{-1}\,({\bar P} + {\bar P}^+ - \Lambda - \Lambda^+)\,G^{-1}\, (\qslash\,G^{-1}\,\qslash^+ + \qslash^+\,G^{-1}\,\qslash)\right]\biggr]\ ,
\label{eq:4.18}
\eea
or, schematically,
\be
\Delta{\cal A}^{(3)} \propto G^{-3}\,[ ({\bar P}- \Lambda)^3 + ({\bar P} - \Lambda)\ \qslash^2]\ ,
\tag{4.18$'$}
\label{eq:4.18'}
\ee
and for the quartic one
\bea
\Delta{\cal A}^{(4)} &=& -2\,\Tr\,\Bigl[ 4\, G^{-1}\,({\bar P}-\Lambda)\,G^{-1}\,
   ({\bar P}-\Lambda)\,G^{-1}\,({\bar P}-\Lambda)\,G^{-1}\,({\bar P}-\Lambda)^+
\nonumber\\
&& \hskip 25pt + 4\, G^{-1}\,({\bar P}-\Lambda)^+\,G^{-1}\,
   ({\bar P}-\Lambda)^+\,G^{-1}\,({\bar P}-\Lambda)^+\,G^{-1}\,({\bar P}-\Lambda)
\nonumber\\
&& \hskip 25pt   + 3\, G^{-1}\,({\bar P}-\Lambda)\,G^{-1}\,
   ({\bar P}-\Lambda)\,G^{-1}\,({\bar P}-\Lambda)^+\,G^{-1}\,({\bar P}-\Lambda)^+
\nonumber\\
&& \hskip 25pt   + 3\, G^{-1}\,({\bar P}-\Lambda)\,G^{-1}\,
   ({\bar P}-\Lambda)^+\,G^{-1}\,({\bar P}-\Lambda)\,G^{-1}\,({\bar P}-\Lambda)^+ \Bigr]
\nonumber\\
&& - 8\,\Tr\,\Bigl[ G^{-1} ({\bar P} +{\bar P}^+ - \Lambda - \Lambda^+)\,G^{-1}\,
   ({\bar P} + {\bar P}^+ - \Lambda -\Lambda^+)\,G^{-1}\,
   \left( \qslash\,G^{-1}\,\qslash^+ + \qslash^+\,G^{-1}\,\qslash\right)
\nonumber\\
&&\hskip 32pt + G^{-1}\,({\bar P} + {\bar P}^+ -\Lambda - \Lambda^+)\,G^{-1}\,
   \qslash\,G^{-1}\,({\bar P} + {\bar P}^+ -\Lambda - \Lambda^+)\,G^{-1}\,
   \qslash^+ \Bigr]
\nonumber\\
&& -4\,\Tr\,\left[ G^{-1}\,\qslash\,G^{-1}\,\qslash^+\,G^{-1}\,\qslash\,G^{-1}\,\qslash^+\right]\ ,
\label{eq:4.19}
\eea
or, schematically,
\be
\Delta{\cal A}^{(4)} \propto G^{-4}\,\left[ ({\bar P}-\Lambda)^4 + ({\bar P}-\Lambda)^2\,\qslash^2
   + \qslash^4\right]\ .
\tag{4.19$'$}
\label{eq:4.19'}
\ee
\end{widetext}
Here we have defined
\be
\qslash = q - \lambda\ ,
\label{eq:4.20}
\ee
and $G^{-1}{\bar P}$ etc. denote matrix products with the matrix elements
of $G$ given by $G_{12} = \delta_{12}\,G_1\,(\tau_0 \otimes s_0)$. The matrix elements of 
${\bar P}$ are ${^i_r {\bar P}}_{n_1n_2}({\bm p}_1,{\bm p}_2)$, see the 
left-hand side of Eq.\ (\ref{eq:2.14c}), and those of ${\bar P}^+$ are 
$\fourchoose{+}{+}{+}{-}{r} \fourchoose{+}{-}{-}{-}{i} {^i_r {\bar P}}_{n_2n_1} (-{\bm p}_2,-{\bm p}_1)$,
see the right-hand side of Eq.\ (\ref{eq:2.14c}). The matrix elements of $\Lambda$, $\Lambda^+$,
$q$, $q^+$, and $\lambda$, $\lambda^+$, respectively, are given by analogous expressions. 
This implies for the expansion of the various matrices in the spin-quaternion basis
\bse
\label{eqs:4.21}
\bea
{\bar P}_{12} &=& \sum_{r=0}^3\sum_{i=0}^3 {^i_r {\bar P}}_{12}\,(\tau_r\otimes s_i)\ ,
\label{eq:4.21a}\\
{\bar P}^+_{12} &=& \sum_{r=0}^3\sum_{i=0}^3 {^i_r {\bar P}}_{21}^{\ddagger}\,
       (\tau_r^+\otimes s_i^+)\ ,
\label{eq:4.21b}
\eea
where 
\be
{^i_r {\bar P}}_{12}^{\ddagger} = {^i_r {\bar P}}_{n_1n_2} (-{\bm p}_1,-{\bm p}_2)
\quad,\quad (r=0,3)
\label{eq:4.21c}
\ee
in the particle-hole channel, and
\be
{^i_r {\bar P}}_{12}^{\ddagger} = - {^i_r {\bar P}}_{n_1n_2} (-{\bm p}_1,-{\bm p}_2)
\quad,\quad (r=1,2)
\label{eq:4.21d}
\ee
in the particle-particle channel.
\ese
$\tau_r^+$ and $s_i^+$ are the hermitian conjugates of $\tau_r$ and $s_i$,
respectively. Analogous relations hold for $\Lambda$, $q$, and $\lambda$ and their adjoints.
In all cases the frequency ordering restrictions explained after Eqs.\ (\ref{eqs:4.16}) apply.

Expressions for the higher-order terms in $\Delta{\cal A}$ in terms of $P$, $q$, $\Lambda$, 
and $\lambda$ can be constructed analogously. For instance, $\Delta{\cal A}^{(5)}$ contains 
terms that have the schematic structures 
\be
\Delta{\cal A}^{(5)} \propto G^{-5}\,[({\bar P}-\Lambda)^5 + ({\bar P}-\Lambda)^3 \qslash^2 
   + ({\bar P}-\Lambda)\,\qslash^4]\ .
\label{eq:4.22}
\ee

\subsection{Integrating out the massive modes}
\label{subsec:IV.C}

We now recall that ${\bar P}$ and $\Lambda$ represent massive fluctuations, whereas $q$ and
$\lambda$ represent soft ones, and our goal is to construct an effective theory in terms of the 
latter. We cannot simply neglect $\bar P$ and $\Lambda$ since they couple to $q$ and $\lambda$; 
rather, we need to integrate them out in an approximation that respects the Ward identity 
discussed in Sec.\ \ref{sec:III}. We let ourselves be guided in this process by an analysis of the 
$O(2)$ $\phi^4$-theory in Appendix \ref{app:B}. The procedure used below for our matrix theory 
is analogous to the one explained for the $O(2)$ model in the context of Eqs.\ (\ref{eqs:B.11}). 
To avoid unnecessary notational complexity, from here on we will restrict ourselves to a model
with only one interaction amplitude, $\Gamma_{\text{s}}^{(0)} \equiv \Gamma$, in the
particle-hole spin-singlet channel. The spin-triplet  ($i=1,2,3$) and particle-particle ($r=1,2$)
degrees of freedom then just represent noninteracting electrons, and in the remainder of this
paper we will neglect them. That is, we keep only matrix elements with $r=0,3$ and $i=0$.
It will be obvious how to generalize the theory to the presence of interactions in other 
channels.\cite{singlet_only_footnote}

\subsubsection{Screening of the interaction}
\label{subsubsec:IV.C.1}

It is advantageous to first eliminate the coupling between ${\bar P}$ and $q$ at the Gaussian level. 
We see from Eq.\ (\ref{eq:4.17}) that this can be achieved by a simple shift of 
${\bar P}$. We define
\bse
\label{eqs:4.23}
\be
{^0_r P}_{12} = {^0_r{\bar P}}_{12} - \frac{2T\Gamma}{V}\sum_{3,4,5,6} M^{-1}_{12,34}\,
   \delta_{3-4,5-6}\,{^0_r q}_{56}\ .
\label{eq:4.23a}
\ee
Here
\bea
M^{-1}_{12,34} &=& \delta_{13}\delta_{24}\Phi_{12} 
\nonumber\\
&& + \delta_{1-2,3-4}\,\frac{2(T/V)\Gamma\,\Phi_{12}\Phi_{34}}{1 - 2T\Gamma        
          \frac{1}{V}\sum_{5,6}\delta_{1-2,5-6} \Phi_{56}}
\nonumber\\
\label{eq:4.23b}
\eea
is the inverse of
\be
M_{12,34} = \delta_{13}\delta_{24}\Phi^{-1}_{12}  - \frac{2T\Gamma}{V}\,\delta_{1-2,3-4}\ .
\label{eq:4.23c}
\ee
Due to the frequency restrictions inherent in $\Phi_{12}$, see Eq.\ (\ref{eq:4.12c}), the 
frequency sum in the denominator of Eq.\ (\ref{eq:4.23b}) has no hydrodynamic content 
and can be replaced by minus the static electron density susceptibility in Hartree-Fock 
approximation, $\chi_{\text{st}}$ which in turn we can replace by $\NF$: 
\be
-\frac{T}{V}\sum_{5,6}\delta_{1-2,5-6}\,\Phi_{56} \approx \chi_{\text{st}} \approx \NF\ .
\label{eq:4.23d}
\ee
\ese
 Equation (\ref{eq:4.23a}) thus simplifies to
\bse
\label{eqs:4.24}
\bea
{^0_r P}_{12} &=& {^0_r{\bar P}}_{12} - 2T\gamma\,
     \Phi_{12}\,\frac{1}{V}\sum_{3,4} \delta_{1-2,3-4}\,{^0_r q}_{34}\qquad
\label{eq:4.24a}\\
&\equiv& {^0_r{\bar P}}_{12} -  {^0_r(} {\hat\Phi}{\hat\gamma} q)_{12}
\label{eq:4.24b}
\eea
with
\be
\gamma = \Gamma/(1 + 2\NF \Gamma)
\label{eq:4.24c}
\ee
the statically screened interaction amplitude and operators
\bea
\hat\gamma _{12,34} &=& (\tau_0\otimes s_0)\,2(T/V)\gamma\,\delta_{1-2,3-4}\ ,
\label{eq:4.24d}\\
\hat\Phi_{12,34} &=& (\tau_0\otimes s_0)\,\delta_{13}\,\delta_{24}\,\Phi_{12}\ .
\label{eq:4.24e}
\eea
\ese
Analogously we define $\hat\Gamma$ and $\hat\varphi$. These operators are all
self-adjoint with respect to the matrix adjoint, i.e.,
$({\hat\gamma} q)^+ = {\hat\gamma} q^+$ etc.

In addition to decoupling $\bar P$ and $q$ in the Gaussian action, this procedure introduces a new term
quadratic in $q$ that combines with the $q^2$ term in Eq.\ (\ref{eq:4.17}) to change the interaction
amplitude $\Gamma$ to $\gamma$. That is, the bilinear coupling between the massive
modes and the soft ones leads to the screening of the interaction. The Gaussian action in terms of
$P$, $\Lambda$, $q$, and $\lambda$ now reads
\begin{widetext}
\bea
{\cal A}^{(2)} &=& -8\sum_{r=0,3} \sum_{1,2\atop 3,4} \left[{^0_r q}_{12}\,\left(\delta_{13}\delta_{24}\,
       \frac{1}{\varphi_{12}}
     - \delta_{1-2,3-4}\,\frac{2T\gamma}{V}\right)\,{^0_r q}_{34}^{\ddagger}
   - \delta_{13}\delta_{24}\,\frac{1}{\varphi_{12}}\,{^0_r\lambda}_{12}\,{^0_r\lambda}_{34}^{\ddagger}\right]
\nonumber\\
&& -8\sum_{r=0,3}\sum_{1,2\atop 3,4} I_{12}\,\left[{^0_r P}_{12}\,
     \left(\delta_{13}\delta_{24}\,\frac{1}{\Phi_{12}} - \delta_{1-2,3-4}\,\frac{2T\Gamma}{V} \right)\,
       {^0_r P}_{34}^{\ddagger} 
     - \delta_{13}\delta_{24}\,\frac{1}{\Phi_{12}}\,{^0_r\Lambda}_{12}\,{^0_r\Lambda}_{34}^{\ddagger}\,\right]\ .
\label{eq:4.25}
\eea
\end{widetext}
The cubic and quartic parts, respectively, of the fluctuation action have the schematic form
\bse
\label{eqs:4.26}
\bea
\Delta{\cal A}^{(3)} &\propto& G^{-3} \bigl[(P - \Lambda + {\hat\Phi}{\hat\gamma} q)^3 
\nonumber\\
&& \hskip 25pt + (P - \Lambda + {\hat\Phi}{\hat\gamma} q) (q-\lambda)^2\bigr]\ ,\qquad
\label{eq:4.26a}\\
\Delta{\cal A}^{(4)} &\propto& G^{-4}\,\bigl[(P - \Lambda + {\hat\Phi}{\hat\gamma} q)^4 
\nonumber\\
&& \hskip 25pt + (P - \Lambda + {\hat\Phi}{\hat\gamma} q)^2 (q-\lambda)^2 
\nonumber\\
&& \hskip 25pt + (q-\lambda)^4\bigr]\ .
\label{eq:4.26b}
\eea
\ese
More explicit expression are obtained by substituting Eqs.\ (\ref{eqs:4.24}) into
Eqs.\ (\ref{eq:4.18}) and (\ref{eq:4.19}).

\subsubsection{Integrating out the massive modes}
\label{subsubsec:IV.C.2}

We now integrate out $P$ and $\Lambda$ in a saddle-point approximation while
keeping $q$ and $\lambda$ fixed. This procedure is analogous to the one explained
in Appendix \ref{app:B} for deriving the $O(2)$ nonlinear sigma model. In the current
context it leads to an effective soft-mode theory that is related to, but has a different
structure than, a sigma model.

The saddle-point equations read
\bse
\label{eqs:4.27}
\bea
0 &=& -16\,I_{12} \left[{^0_r (}{\hat\Phi}^{-1}P)_{12}^{\ddagger} 
                 - {^0_r({\hat\Gamma} P)}_{12}^{\ddagger}\right] + \delta\Delta{\cal A}/\delta {^0_r P}_{12}\ ,
\nonumber\\
\label{eq:4.27a}\\
0 &=& 16\,I_{12}\  {^0_r({\hat\Phi}^{-1} \Lambda)}_{12}^{\ddagger} 
                 - \delta\Delta{\cal A}/\delta {^0_r P}_{12}\ .
\label{eq:4.27b}
\eea
\ese
Here we have used $\delta\Delta{\cal A}/\delta P = -\delta\Delta{\cal A}/\delta\Lambda$,
and ${\hat\Gamma} P$ is defined in analogy to ${\hat\gamma} q$ in Eq.\ (\ref{eq:4.24b}). From Eqs.
(\ref{eqs:4.27}) we immediately obtain the useful identities
\bea
(P - \Lambda)_{12} &=& ({\hat\Phi}{\hat\Gamma} P)_{12}
\nonumber\\
&=& ({\hat\Phi}{\hat\gamma}\Lambda)_{12}\ .
\label{eq:4.28}
\eea
$P$ is thus given in terms of $\Lambda$, which in turn can be expressed in terms of $q$
and $\lambda$ by solving Eq.\ (\ref{eq:4.27b}). We will determine the action explicitly to
$O(q^4)$,\cite{powers_of_q_footnote} which requires $\Lambda$ to $O(q^2)$. 
From Eqs.\ (\ref{eq:4.27b}) and (\ref{eq:4.26a}) (or, more explicitly, from Eqs. (\ref{eq:4.18})
and (\ref{eqs:4.24}); note that only $\Delta{\cal A}^{(3)}$ contributes to $\Lambda$ to this
order) we find
\bea
\Lambda_{12} &=& -2i(\qslash\,G^{-1}\,\qslash^+ + \qslash^+\,G^{-1}\,\qslash)_{12}
\nonumber\\
&& - 2i\,\left[(({\hat\Phi}{\hat\gamma} q)\, G^{-1}\, ({\hat\Phi}{\hat\gamma} q^+))_{12} \right.
\nonumber\\
&& \hskip 30pt   + (({\hat\Phi}{\hat\gamma} q^+)\, G^{-1}\, ({\hat\Phi}{\hat\gamma} q))_{12}
\nonumber\\
&& \hskip 40pt  \left.   + (({\hat\Phi}{\hat\gamma} q)\, G^{-1}\, ({\hat\Phi}{\hat\gamma} q))_{12}\right] + O(q^3).
\nonumber\\
\label{eq:4.29}
\eea
Terms of higher order in $q$ can be obtained by means of an obvious iteration procedure.

Fluctuations about this saddle point yield measure terms that reflect the Jacobian due to
the change from integrating over $\delta Q$ and $\delta{\tilde\Lambda}$ to obtain the
partition function to integrating over $q$ and $\lambda$.  For the $O(2)$ nonlinear 
sigma model this is demonstrated in Appendix \ref{app:B}. While they are structurally
important for preserving the symmetry of the action, these terms are not of
practical importance for a loop expansion near the lower critical dimensionality
$d_{\text c}^-$ of the problem. The reason is that they lead to loop integrals
that are less infrared divergent than those that result from the terms we have derived so far. 
In the $O(2)$
nonlinear sigma model they carry no gradients, see Eq.\ (\ref{eq:B.12}), and hence are
less infrared divergent near $d_{\text c}^- = 2$ by two degrees than the terms contained
in the nonlinear sigma model without the measure terms. The same statement is true for
the generalized nonlinear sigma model that describes the Anderson-Mott metal-insulator
transition in the disordered interacting fermion problem, which also has
$d_{\text c}^- = 2$.\cite{Finkelstein_1983, Belitz_Kirkpatrick_1994} For the current
problem the lower critical dimension is $d_{\text c}^- = 1$, the dimension at and
below which the Fermi liquid is not stable for any parameter values. This reduction of
$d_{\text c}^-$ compared to the disordered case is a consequence of the soft modes
having a ballistic dispersion relation, as opposed to a diffusive one. For the same
reason the measure terms will lead to loop integrals that, near $d_{\text c}^- = 1$,
are less divergent by one power of momentum than the ones we keep. For the sake
of simplicity we will ignore these measure terms in what follows; it is clear how to
derive them if that's desirable.

\subsection{Effective action}
\label{subsec:IV.D}

\subsubsection{Effective action to $O(q^4)$}
\label{subsubsec:IV.D.1}

We are now in a position to write the effective action in terms of the soft modes $q$ and 
$\lambda$ only. The relation between ${\hat\Gamma} P$ and ${\hat\gamma}\Lambda$ 
in Eq.\ (\ref{eq:4.28}) allows to write the Gaussian action, Eq.\ (\ref{eq:4.25}),
\begin{widetext}
\be
{\cal A}^{(2)} = -8\sum_{r=0,3} \sum_{1,2\atop 3,4} \left[{^0_r q}_{12}\,\left(\delta_{13}\delta_{24}\,
   \frac{1}{\varphi_{12}}
     - \delta_{1-2,3-4}\,\frac{2T\gamma}{V}\right)\,{^0_r q}_{34}^{\ddagger}
   - \delta_{13}\delta_{24}\,\frac{1}{\varphi_{12}}\,{^0_r\lambda}_{12}\,{^0_r\lambda}_{34}^{\ddagger}
   + {^0_r \Lambda}_{12}\,\frac{2T\gamma}{V}\,\delta_{1-2,3-4}\,{^0_r \Lambda}_{34}^{\ddagger} \right]\ .
\label{eq:4.30}
\ee
\end{widetext}
This can be written more compactly
\bea
{\cal A}^{(2)} &=& -2\,\Tr [q({\hat\varphi}^{-1} - {\hat\gamma})q^+] +2\,\Tr\,(\lambda\,{\hat\varphi}^{-1}\,\lambda^+)
\nonumber\\
&&   - 2\,\Tr\,(\Lambda\,{\hat\gamma}\,\Lambda^+)\ .
\label{eq:4.31}
\eea
Here $\Lambda$ is given
in terms of $q$ and $\lambda$ by Eq.\ (\ref{eq:4.29}). Similarly, the other terms in 
the action, $\Delta{\cal A}^{(3)}$, $\Delta{\cal A}^{(4)}$, etc., can be expressed in terms of
$q$ and $\lambda$ by means of Eqs.\ (\ref{eqs:4.24}), (\ref{eq:4.28}), and (\ref{eq:4.29}).
The resulting action is the desired effective action in terms of the soft modes $q$ and
$\lambda$. 

Note that this is a local field theory, i.e., all vertices are finite in the limit
of small wave numbers and small frequencies. This can be seen most easily from
Eqs.\ (\ref{eqs:4.26}) and their generalizations: The terms with the highest power of
Green functions in $\Delta{\cal A}^{(n)}$ has the structure 
$G^{-n}\Phi^n(\gamma q)^n \sim G^n (\gamma q)^n$.
Since $G \sim 1/\Omega$ and $\gamma q \sim \Omega q$, this vertex is of $O(1)$,
and all other vertices scale as a positive power of the frequency.

One might consider analyzing this theory by means of a loop expansion in terms of
both $q$ and $\lambda$. However, this is not desirable. The reason is that the origin
of $\lambda$ is a hard (i.e., delta-function) constraint that constrains the $q$ to soft
bilinear fermion modes. Treating $\lambda$ perturbatively relaxes this hard constraint,
which effectively introduces spurious soft modes. (We note that no such difficulty arises
for $\Lambda$, which enforces a constraint for massive modes.) It therefore is desirable
to treat $\lambda$ exactly. This can be achieved by 
eliminating $\lambda$ in favor of diagram rules for an action that is formulated 
entirely in terms of $q$ and $\qslash$. To see this let us consider the terms quadratic
in $q$ and $\lambda$, i.e., the first two terms in Eq.\ (\ref{eq:4.30}) or (\ref{eq:4.31}).
The Gaussian $q$ and $\lambda$-propagators are obtained by inverting the corresponding
vertices. We see that the $\lambda$-propagator is minus the noninteracting part of the 
$q$-propagator, which is given by $\varphi$, and that $q$ and
$\lambda$ are not coupled in the Gaussian action. This implies that
whenever $\qslash$ is contracted with $\qslash$ , the resulting
propagator is effectively equal to the interacting part of the $q$-propagator,
whereas contracting $\qslash$ with $\lambda$ results in the full
$q$-propagator, as does contracting $q$ with
$q$.\cite{diagram_rules_footnote} $\lambda$ can thus formally be integrated
out, despite the fact that it is a soft fluctuation, without causing undesirable
features of the effective theory, such as nonlocality. The reason is that the
{\em only} effect of $\lambda$ is to cancel well-defined contributions from
other soft fluctuations. We thus arrive at the following effective soft-mode
action:
\bse
\label{eqs:4.32}
\be
{\cal A}_{\text{eff}} = {\cal A}_{\text{eff}}^{(2)} + \Delta{\cal A}_{\text{eff}}^{(3)}
+ \Delta{\cal A}_{\text{eff}}^{(4)} + O(q^5)\ .
\label{eq:4.32a}
\ee
with a Gaussian part
\begin{widetext}
\bea
{\cal A}_{\text{eff}}^{(2)} &=& -2\,\Tr\,[q({\hat\varphi}^{-1} - {\hat\gamma})q^+]
\nonumber\\
&\equiv& -8\sum_{r=0,3} \sum_{1,2\atop 3,4} {^0_r q}_{12}\,
     \left(\delta_{13}\delta_{24}\,\frac{1}{\varphi_{12}}
     - \delta_{1-2,3-4}\,\frac{2T\gamma}{V}\right)\,{^0_r q}_{34}^{\ddagger}
\nonumber\\
\label{eq:4.32b}
\eea
and non-Gaussian parts $\Delta{\cal A}_{\text{eff}}^{(n)}$ that each
contain terms of $O(q^n)$. The first two we obtain from
Eqs.\ (\ref{eq:4.18}), (\ref{eq:4.19}), and (\ref{eq:4.24b}) as follows,\cite{frequency_restrictions_footnote}
\bea
\Delta{\cal A}_{\text{eff}}^{(3)} &=& -4i\,\Tr\,\left[ G^{-1}\,({\hat\Phi}{\hat\gamma} q + {\hat\Phi}{\hat\gamma} q^+)\,G^{-1}\,
   (\qslash\,G^{-1}\,\qslash^+ + \qslash^+\,G^{-1}\,\qslash) \right]
\nonumber\\
&& -4i\,\Tr\,\left[ G^{-1}({\hat\Phi}{\hat\gamma}\,q)\,G^{-1}\,({\hat\Phi}{\hat\gamma} q + {\hat\Phi}{\hat\gamma} q^+)\,G^{-1}\,({\hat\Phi}{\hat\gamma}\,q^+)\right]\ ,
\nonumber\\
\label{eq:4.32c}
\eea
\bea
\Delta{\cal A}_{\text{eff}}^{(4)} = -4&&\hskip -10pt\Tr\left[G^{-1}\,\qslash\, G^{-1}\, \qslash^+\, G^{-1}\, \qslash\, G^{-1}\, \qslash^+\right]
\nonumber\\
 -4i&&\hskip -10pt\Tr\left[ G^{-1}\,({\hat\Phi}{\hat\gamma}\Lambda + {\hat\Phi}{\hat\gamma}\Lambda^+) \,G^{-1}\,
   (\qslash\,G^{-1}\,\qslash^+ + \qslash^+\,G^{-1}\,\qslash) \right] 
\nonumber\\
- 2&& \hskip -10pt \Tr\,(\Lambda{\hat\gamma}\Lambda^+)
\nonumber\\
-8&&\hskip -10pt\Tr\,\Bigl[ G^{-1}({\hat\Phi}{\hat\gamma} q + {\hat\Phi}{\hat\gamma} q^+)\,G^{-1}\,
     ({\hat\Phi}{\hat\gamma} q + {\hat\Phi}{\hat\gamma} q^+)\,G^{-1}\, (\qslash\,G^{-1}\,\qslash^+ + \qslash^+\,G^{-1}\,\qslash)
\nonumber\\
&& \hskip 100pt   + G^{-1}\,({\hat\Phi}{\hat\gamma} q + {\hat\Phi}{\hat\gamma} q^+)\,G^{-1}\,\qslash\,G^{-1}\,({\hat\Phi}{\hat\gamma} q + {\hat\Phi}{\hat\gamma} q^+) 
     \,G^{-1}\,\qslash^+ \Bigr]
\nonumber\\
-4i&&\hskip -10pt\Tr\,\left[ G^{-1}\,({\hat\Phi}{\hat\gamma}\Lambda)\,G^{-1}\,({\hat\Phi}{\hat\gamma} q + {\hat\Phi}{\hat\gamma} q^+)\,G^{-1}\,({\hat\Phi}{\hat\gamma} q^+) 
   + G^{-1}\,({\hat\Phi}{\hat\gamma} q)\,G^{-1}\,({\hat\Phi}{\hat\gamma}\Lambda + {\hat\Phi}{\hat\gamma}\Lambda^+)\,G^{-1}\,({\hat\Phi}{\hat\gamma} q^+)  \right.
\nonumber\\
&& \hskip 10pt \left. + G^{-1}\,({\hat\Phi}{\hat\gamma} q)\,G^{-1}\,({\hat\Phi}{\hat\gamma} q + {\hat\Phi}{\hat\gamma} q^+)\,G^{-1}\,({\hat\Phi}{\gamma}\Lambda^+)\right]
\nonumber\\
-2&&\hskip -10pt\Tr\,\Bigl[ 4\,G^{-1}\, ({\hat\Phi}{\hat\gamma} q)\,G^{-1}\,({\hat\Phi}{\hat\gamma} q)\,G^{-1}\,({\hat\Phi}{\hat\gamma} q)\,G^{-1}\,({\hat\Phi}{\hat\gamma} q^+)
                       + 4\,G^{-1}\, ({\hat\Phi}{\hat\gamma} q^+)\,G^{-1}\,({\hat\Phi}{\hat\gamma} q^+)\,G^{-1}\,({\hat\Phi}{\hat\gamma} q^+)\,G^{-1}\,({\hat\Phi}{\hat\gamma} q)
\nonumber\\
&& \hskip 10pt      + 3\,G^{-1}\,({\hat\Phi}{\hat\gamma} q^+)\,G^{-1}\,({\hat\Phi}{\hat\gamma} q^+)\,G^{-1}\,({\hat\Phi}{\hat\gamma} q)\,G^{-1}\,({\hat\Phi}{\hat\gamma} q)
\nonumber\\
&& \hskip 10pt   + 3\,G^{-1}\,({\hat\Phi}{\hat\gamma} q^+)\,G^{-1}\,({\hat\Phi}{\hat\gamma} q)\,G^{-1}\,({\hat\Phi}{\hat\gamma} q^+)\,G^{-1}\,({\hat\Phi}{\hat\gamma} q) \Bigr]\ .
\nonumber\\
\label{eq:4.32d}
\eea
Here $\Lambda$ is given by Eq.\ (\ref{eq:4.29}), which we restate for completeness:
\bea
\Lambda &=& -2i(\qslash\,G^{-1}\,\qslash^+ + \qslash^+\,G^{-1}\,\qslash)
\nonumber\\
&& -2i\left[({\hat\Phi}{\hat\gamma} q)\,G^{-1}\,({\hat\Phi}{\hat\gamma} q^+) + ({\hat\Phi}{\hat\gamma} q^+)\,G^{-1}\,({\hat\Phi}{\hat\gamma} q)
   + ({\hat\Phi}{\hat\gamma} q^+)\,G^{-1}\,({\hat\Phi}{\hat\gamma} q^+)\right]\ .
\label{eq:4.32e}
\eea
\end{widetext}
\ese
For the effective action to $O(q^4)$, $\Lambda$ is needed only to $O(q^2)$.

This action needs to be augmented by a prescription for the role of $\qslash$. 
From the above discussion it follows that contractions of $\qslash$ with $q$
are given by the $q$-propagator as it follows from ${\cal A}_{\text{eff}}^{(2)}$,
\bse
\label{eqs:4.33}
\bea
\langle {^0_r q}_{12}\,{^0_s q}^{\ddagger}_{34}\rangle &=& 
   \langle {^0_r \qslash}_{12}\,{^0_s q}^{\ddagger}_{34}\rangle 
   = \langle {^0_r q}_{12}\,{^0_s \qslash}^{\ddagger}_{34}\rangle =
\nonumber\\
&& \hskip -37pt = \frac{1}{16}\,\delta_{rs}\,
   \Bigl[\delta_{13}\delta_{24}\,\varphi_{12} + \frac{2\gamma T}{V}\delta_{1-2,3-4}\,
   \frac{\varphi_{12}\,\varphi_{34}}{1 + 2\gamma\chi^{(0)}_{1-2}}\Bigr]\ ,
\nonumber\\
\label{eq:4.33a}
\eea
while contractions of $\qslash$ with $\qslash$ are given by
\be
\langle {^0_r \qslash}_{12}\,{^0_s \qslash}^{\ddagger}_{34}\rangle =
   \frac{\gamma T}{8V}\, \delta_{1-2,3-4}\,
   \frac{\varphi_{12}\,\varphi_{34}}{1 + 2\gamma\chi^{(0)}_{1-2}}\ .
\ee
\label{eq:4.33b}
\ese
Here we have defined
\be
\chi^{(0)}_{1-2} = -\frac{T}{V}\sum_{34} \delta_{1-2,3-4}\,\varphi_{34}\ .
\label{eq:4.34}
\ee
Physically, $\chi^{(0)}$ is the hydrodynamic part of the density susceptibility 
per spin in Hartree-Fock approximation, see the discussion below.

Equations (\ref{eqs:4.32}) through (\ref{eq:4.34}) completely specify the effective
theory for the purposes of a loop expansion for $q$-correlation 
functions.\cite{vertex_functions_footnote}
They are the central formal result of the present paper.
Their derivation makes is clear how to derive the effective action to any
desired power in $q$.

\subsubsection{Gaussian propagators}
\label{subsubsec:IV.D.2}

In order to make the theory suitable for actual calculations we need to
explicitly evaluate the correlation function $\chi^{(0)}$ defined in Eq.\ 
(\ref{eq:4.34}). If we absorb the Hartree-Fock self energy into the
chemical potential, as explained in Sec.\ \ref{subsubsec:IV.B.1}, the
integral that defines $\chi^{(0)}$ reads explicitly
\bea
\chi^{(0)}({\bm k},i\Omega_n) &=& -T\sum_{m=0}^{n-1} \frac{1}{V}\sum_{\bm p}
   \frac{1}{i\omega_m - \xi_{{\bm p}+{\bm k}/2}} \,
\nonumber\\
&& \times\frac{1}{i\omega_m - i\Omega_n
     - \xi_{{\bm p}-{\bm k}/2}}\ .
\label{eq:4.35}
\eea
The integral can be performed exactly. However, this is not necessary
for most purposes, as the crucial hydrodynamic structure of $\chi^{(0)}$
is preserved in the well-known approximation, valid for small values of
$k \equiv \vert{\bm k}\vert$ and $\vert\Omega_n\vert$, that performs the
radial part of the momentum integral by means of a contour integration
over the interval $-\infty < \xi_{\bm p} < \infty$, which makes the
frequency sum trivial.\cite{Abrikosov_Gorkov_Dzyaloshinski_1963}
Within this scheme, which we refer to as the AGD approximation, we
are left with an angular integral only,
\bse
\label{eqs:4.36}
\be
\chi^{(0)}({\bm k},i\Omega_n) = -\NF\,\frac{G\Omega_n}{k}\,\varphi_d(G i\Omega_n/k)\ ,
\label{eq:4.36a}
\ee
where
\be
\varphi_d(z) = \frac{-i}{S_{d-1}} \int d\Omega_{\bm p}\,\frac{1}{\hat{\bm p}\cdot\hat{\bm k} - z}
\label{eq:4.36b}
\ee
is a causal function of the complex frequency $z$. $G$ is a coupling constant
whose bare value is 
\be
G = 1/\vF\ .
\label{eq:4.36c}
\ee
\ese
It constitutes the natural coupling constant for a loop expansion.
$S_{d-1} = 2\pi^{d/2}/\Gamma(d/2)$
with $\Gamma(x)$ the Gamma function is the
surface area of the unit $(d-1)$-sphere, and $d\Omega_{\bm p}$ is the
angular integration measure for the unit vector $\hat{\bm p}$ for fixed unit
vector $\hat{\bm k}$  in $d$ dimensions. For general dimensions $d>1$
the latter takes the form\cite{Stillinger_1977, He_1991}
\bse
\label{eqs:4.37}
\be
\int d\Omega_{\bm p} = \frac{2\pi^{\epsilon/2}}{\Gamma(\epsilon/2)}
   \int_{-1}^{1} \frac{d\eta}{(1-\eta^2)^{1-\epsilon/2}}\ ,\  (\epsilon = d-1 > 0)\ ,
\label{eq:4.37a}
\ee
where $\epsilon = d-1$ and
$\eta = \cos\theta$ with $\theta = \sphericalangle (\hat{\bm p},\hat{\bm k})$.
The limiting expression for $d\to 1$ is
\be
\int d\Omega_{\bm p} = \int_{-\infty}^{\infty} d\eta\ [\delta(\eta + 1) + \delta(\eta - 1)]
   \quad,\quad (d=1)\ .
\label{eq:4.37b}
\ee
\ese
From Eqs.\ (\ref{eq:4.36b}), (\ref{eqs:4.37}) we obtain
\be
\varphi_d(z) = \frac{i}{z}\,{_2F}_1(1,1/2,d/2;1/z^2)\ ,
\label{eq:4.38}
\ee
where ${_2F}_1$ is Gauss's hypergeometric function. In $d=1,2,3$ this reduces to the
familiar expressions for the hydrodynamic part of the Lindhard function in these
dimensions:
\bse
\label{eqs:4.39}
\bea
\varphi_{d=1}(z) &=& -iz/(1-z^2)\ ,
\label{eq:4.39a}\\
\varphi_{d=2}(z) &=& \sgn(\Im z)/\sqrt{1-z^2}
\nonumber\\
                   &\equiv& i/\sqrt{z+1}\sqrt{z-1}\ ,
\label{eq:4.39b}\\
\varphi_{d=3}(z) &=& \frac{-i}{2}\,\ln\left(\frac{1-z}{-1-z}\right)\ .
\label{eq:4.39c}
\eea
\ese

The Gaussian propagators, Eqs.\ (\ref{eqs:4.33}), are now explicitly specified.

\subsubsection{The Goldstone propagator and the density susceptibility}
\label{subsubsec:IV.D.3}

\begin{figure}[t]
\vskip -0mm
\includegraphics[width=8.0cm]{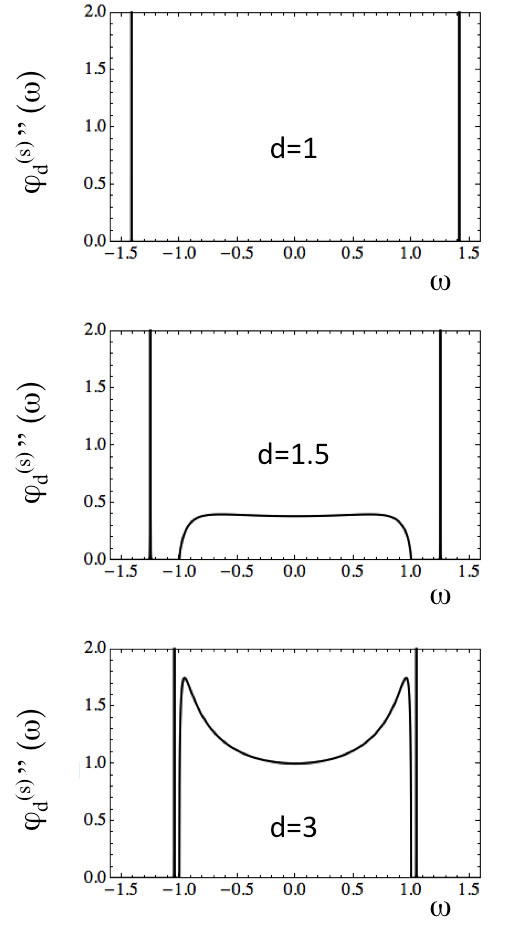}
\caption{The spectrum $\varphi^{({\text s})\prime\prime}_d(\omega) 
              = \Re \varphi^{({\text s})}_d(z=\omega + i0)$ for $2\NF\gamma = 0.5$ 
              as a function of the frequency $\omega$ in dimensions $d=1$, $d=1.5$, and $d=3$.
              The vertical lines denote $\delta$-function contributions to the spectrum.}
\label{fig:4.1}
\end{figure}

From the preceding two subsections we see that the Goldstone modes are
given by the propagator $\varphi_d(z)$, Eq.\ (\ref{eq:4.38}), which determines
the density susceptibility in saddle-point (i.e., Hartree-Fock) approximation via 
Eq.\ (\ref{eq:4.36a}). This is just the hydrodynamic part of the familiar Lindhard 
function, see Eqs.\ (\ref{eqs:4.39}). It is illustrative to also consider the density 
susceptibility in the Gaussian approximation to the effective field theory. From
Eq.\ (\ref{eq:2.18a}) we see that the density susceptibility $\chi$ is given by a
$\langle q\,q\rangle$ correlation function plus terms that result from the
coupling of $q$ to ${\bar P}$. Eliminating that coupling by the shift that
was discussed in Sec.\ \ref{subsubsec:IV.C.1} leads to Fermi-liquid corrections.
Using Eq.\ (\ref{eq:4.33a}) we obtain for the density susceptibility in Gaussian
approximation
\be
\chi({\bm k},i\Omega_n) = \frac{1}{(1+2\NF\Gamma)^2}\,
   \frac{2\chi^{(0)}({\bm k},i\Omega_n)}{1 + 2\gamma \chi^{(0)}({\bm k},i\Omega_n)}\ .
\label{eq:4.40}
\ee
We recognize this as the random-phase approximation (RPA) for the density susceptibility 
\cite{Pines_Nozieres_1989} with additional Fermi-liquid corrections. One of its 
characteristic features is the collective mode known as zero sound that results from  
a real root of the denominator. To see this we define another causal function
\be
\varphi^{({\text s})}_d(z) = \frac{1}{\varphi_d^{-1}(z) + i2\NF\gamma z}\ .
\label{eq:4.41}
\ee
$\chi$ can be expressed in terms of this as
\be
\chi({\bm k},i\Omega_n) = \frac{-2\NF}{(1+2\NF\Gamma)^2}\,\frac{G\Omega_n}{k}\,\varphi^{({\text s})}_d(Gi\Omega_n/k)\ .
\label{eq:4.42}
\ee
From Eq.\ (\ref{eq:4.36a}) we see that, apart from the Fermi-liquid corrections, 
$\varphi^{({\text s})}_d$ relates to $\varphi_d$
the same way $\chi$ relates to $\chi^{(0)}$. Using Eqs.\ (\ref{eq:4.38}, \ref{eqs:4.39}) in
Eq.\ (\ref{eq:4.41}) one finds that in $d=1$ the spectrum of $\varphi^{({\text s})}_d$ is
exhausted by the zero-sound poles. In $d>1$ a particle-hole continuum emerges, but
the zero-sound poles outside of the continuum remain. This is illustrated in Fig.\ \ref{fig:4.1}.
Note that the zero-sound modes are collective density fluctuations that are the
result of particle-number conservation, whereas the Goldstone modes result from 
the spontaneously broken continuous symmetry discussed in Sec.\ \ref{sec:III} and
are {\em not} related to a conservation law. This manifests itself, for instance, in
the fact that the Goldstone modes acquire a mass at nonzero temperatures, whereas
the zero-sound mode does not.

Equation (\ref{eq:4.42}) provides an important identity that allows to identify the 
thermodynamic density susceptibility $\partial n/\partial\mu$ in terms of the
parameters of the field theory. For simplicity we consider the case $d=1$. Using
Eqs.\ (\ref{eq:4.39a}) and (\ref{eq:4.41}) in (\ref{eq:4.42}) yields
\bse
\label{eqs:4.43}
\bea
\chi({\bm k},i\Omega_n) &=& \frac{-2\NF}{1+2\NF\Gamma}\,\frac{\Omega^2}{(1+2\NF\Gamma)k^2/G^2 + \Omega^2}
\nonumber\\
&&\hskip 90pt (d=1)\ .
\label{eq:4.43a}
\eea
The compressibility sum rule,\cite{Pines_Nozieres_1989} 
\be
\partial n/\partial\mu = - \chi({\bm k}=0,i\Omega_n)
\label{eq:4.43b}
\ee
\ese
then yields
\be
\frac{\partial n}{\partial\mu} = \frac{2\NF}{1+2\NF\Gamma}\ .
\label{eq:4.44}
\ee
This is consistent with another argument. $\partial n/\partial\mu$ is related
to the zero-sound speed $s$ by $s^2 = (n/m)/(\partial n/\partial\mu)$.
\cite{Abrikosov_Gorkov_Dzyaloshinski_1963} From Eq. (\ref{eq:4.43a}) we see
that $s^2 = (1+2\NF\Gamma)/G^2$, which also yields Eq.\ (\ref{eq:4.44}).

While we have derived this relation from the Gaussian theory, it is expected to
be an exact identity that remains true if $\partial n/\partial\mu$ and $\Gamma$ are replaced
by their renormalized counterparts. This is because the structure of the density susceptibility,
which is governed by particle-number conservation, must stay the same.
This is an important point since it is known from perturbation theory that
$\partial n/\partial\mu$ is not singularly renormalized near $d=1$.
\cite{Belitz_Kirkpatrick_Vojta_1997}

\subsection{Density formulation of the effective field theory}
\label{subsec:IV.E}

As we have mentioned after Eq.\ (\ref{eq:4.15}) and (\ref{eq:4.31}), the current theory 
explicitly keeps all of the soft modes that were identified by the Ward identity in 
Sec.\ \ref{sec:III} and hence is local. This is in contrast to the theory formulated in
Ref.\ \onlinecite{Belitz_Kirkpatrick_1997}, which was a formulation in terms of 
s-wave ($\ell=0$) modes only. Technically this is achieved by constraining 
$B_{nm}({\bm x},{\bm x})$, whose spatial Fourier transform is given by Eq.\ (\ref{eq:2.10a}), 
to a local matrix $Q_{nm}({\bm x})$
by means of a Lagrange multiplier field ${\tilde\Lambda}_{nm}({\bm x})$, and
formulating the theory in terms of $Q({\bm x})$ and ${\tilde\Lambda}({\bm x})$. 
This amounts to integrating out the modes in the higher angular momentum channels, and
since these are soft the result is a nonlocal field theory. While this is undesirable for
many purposes, for certain calculations it may be advantageous since it makes it easier
to see which loop integrals as infrared divergent and which are not. The nonlocality,
on the other hand, makes no difference in practice unless one is explicitly interested
in angular momentum channels with $\ell \geq 1$ or wishes to calculate non-local
correlation functions. In the remainder of the section we therefore present the theory
in such a density formulation.

The Fermi-liquid saddle-point solution is the same as in Sec.\ \ref{subsubsec:IV.A.1},
except that the saddle-point value of $Q$ is a position-independent object $Q_n$ given
by Eq.\ (\ref{eq:4.3a}) summed over the wave vector ${\bm k}$. The further development
proceeds as in the phase-space formulation, except that the various products of
Green functions or their inverses get replaced by products of convolutions of Green
functions in wave-vector space and inverses of such convolutions. The frequency
structure of the theory remains unchanged. The Gaussian action, after decoupling ${\bar P}$
and $q$, then reads
\begin{widetext}
\bea
{\cal A}^{(2)} &=& -8 \sum_{r=0,3} \sum_{1,2\atop 3,4} \frac{1}{V}\sum_{\bm k} \left[{^0_r q}_{12}({\bm k})
   \left(\delta_{13}\,\delta_{24}\,\frac{1}{\varphi_{12}({\bm k})} - \delta_{1-2,3-4} 2T\gamma
   \right)\,{^0_r q}_{34}(-{\bm k}) - \delta_{13}\,\delta_{24}\,\frac{1}{\varphi_{12}({\bm k})}\,
   {^0_r\lambda}_{12}({\bm k}){^0_r\lambda}_{34}(-{\bm k})\right]
\nonumber\\
&& \hskip -20pt-8 \sum_{r=0,3} \sum_{1,2\atop 3,4} I_{12}\frac{1}{V}\sum_{\bm k} \left[{^0_r P}_{12}({\bm k})
   \left(\delta_{13}\,\delta_{24}\,\frac{1}{\Phi_{12}({\bm k})} - \delta_{1-2,3-4} 2T\Gamma
   \right)\,{^0_r P}_{34}(-{\bm k}) - \delta_{13}\,\delta_{24}\,\frac{1}{\Phi_{12}({\bm k})}\,
   {^0_r\Lambda}_{12}({\bm k}){^0_r\Lambda}_{34}(-{\bm k})\right]\ .
\nonumber\\
\label{eq:4.45}
\eea
This replaces Eq.\ (\ref{eq:4.25}). In this subsection, $1 \equiv n_1$, etc. represents Matusbara
frequency labels only, in contrast to the notation used in the other parts of this paper.
We have defined
\bse
\label{eqs:4.46}
\bea
\phi_{12}({\bm k}) &=& \frac{1}{V}\sum_{\bm p} G_{n_1}({\bm p})\,G_{n_2}({\bm p}-{\bm k})
\nonumber\\
&\equiv& \begin{cases} \Phi_{12}({\bm k}) & \text{if $(n_1+1/2)(n_2+1/2) > 0$}\\
                              \varphi_{12}({\bm k}) & \text{if $(n_1+1/2)(n_2+1/2) < 0$}\ ,
        \end{cases}
\nonumber\\
\label{eq:4.46a}
\eea
which replaces Eq.\ (\ref{eq:4.12c}). Similarly, the products of inverse Green functions in the higher
terms in the fluctuation expansion get replaced by more complex expressions, which involve
convolutions of Green functions, with the same net power of Green functions. To account for this,
we define
\bea
\phi^{(m)}_{1\ldots m}({\bm k}_1, \ldots , {\bm k}_m) &=& \frac{1}{V}\sum_{\bm p} G_{n_1}({\bm p})\,G_{n_2}({\bm p}-{\bm k}_1-{\bm k}_2)
       \cdots G_{n_m}({\bm p}-{\bm k}_1 - \ldots - {\bm k}_m)
\nonumber\\
&\equiv& \begin{cases} \Phi^{(m)}_{1\ldots m}({\bm k}_1, \ldots , {\bm k}_m) & \text{if $n_1$ through $n_m$ all have the same sign}\\
                                      \varphi^{(m)}_{1\ldots m}({\bm k}_1, \ldots , {\bm k}_m) & \text{if $n_1$ through $n_m$ do not all have the same sign}\ ,
                \end{cases}
\label{eq:4.46b}
\eea
\ese
The cubic and quartic terms,  instead of Eqs.\ (\ref{eq:4.18}, \ref{eq:4.19}),
now take the form
\bse
\label{eqs:4.47}
\bea
\Delta{\cal A}^{(3)} &=& \frac{-4i}{V^{2}} \sum_{1,2,3} \sum_{{\bm k}_1,{\bm k}_2,{\bm k}_3} \delta_{{\bm k}_1+{\bm k}_2+{\bm k}_3,0}\,
\nonumber\\
&&\times\Bigl\{ \Phi^{(3)}_{123}({\bm k}_1,{\bm k}_2)\,\Phi^{-1}_{12}({\bm k}_1)\,\Phi^{-1}_{23}({\bm k}_2)\,
   \Phi^{-1}_{31}({\bm k}_3)\, \Bigl\{\tr\left[({\bar P} - \Lambda)_{12}({\bm k}_1)({\bar P} + {\bar P}^+ 
   - \Lambda - \Lambda^+)_{23}({\bm k}_2) ({\bar P }^+ - \Lambda^+)_{31}({\bm k}_3) \right]
\nonumber\\
&&\hskip -10pt + \varphi^{(3)}_{123}({\bm k}_1,{\bm k}_2)\,\Phi^{-1}_{12}({\bm k}_1)\,\varphi^{-1}_{23}({\bm k}_2)\,
   \varphi^{-1}_{31}({\bm k}_3)\,\tr\left[({\bar P} + {\bar P}^+ - \Lambda - \Lambda^+)_{12}
   ({\bm k}_1)\bigl(\qslash_{23}({\bm k}_2)\qslash^+_{31}({\bm k}_3)
   + \qslash^+_{23}({\bm k}_2)\qslash_{31}({\bm k}_3)\bigr)\right]\Bigr\}\ .
\nonumber\\
\label{eq:4.47a}
\eea
\be
\Delta{\cal A}^{(4)} = \Delta{\cal A}^{(4,0)} + \Delta{\cal A}^{(2,2)} + \Delta{\cal A}^{(0,4)}\ ,
\label{eq:4.47b}
\ee
with
\bea
\Delta{\cal A}^{(4,0)} &=& \frac{-2}{V^3} \sum_{1,2,3,4} \sum_{{\bm k}_1,{\bm k}_2,{\bm k}_3,{\bm k}_4}
   \delta_{{\bm k}_1+{\bm k}_2+{\bm k}_3+{\bm k}_4,0}\,
      \Phi^{(4)}_{1234}({\bm k}_1,{\bm k}_2,{\bm k}_3)\,
        \Phi^{-1}_{12}({\bm k}_1)\,\Phi^{-1}_{23}({\bm k}_2)\,
          \Phi^{-1}_{34}({\bm k}_3)\,\Phi^{-1}_{41}({\bm k}_4)
\nonumber\\
&&\times \Bigl[4\,\tr\left[({\bar P} - \Lambda)_{12}({\bm k_1}) ({\bar P} - \Lambda)_{23}({\bm k_2}) ({\bar P} - \Lambda)_{34}({\bm k_3}) ({\bar P} - \Lambda)^+_{41}({\bm k_4})\right]
\nonumber\\
&&\hskip 10pt + 4\,\tr\left[({\bar P} - \Lambda)^+_{12}({\bm k_1})
    ({\bar P} - \Lambda)^+_{23}({\bm k_2}) ({\bar P} - \Lambda)_{34}^+({\bm k_3}) 
    ({\bar P} - \Lambda)_{41}({\bm k_4})\right]
\nonumber\\
&& \hskip 10pt +3\,\tr\left[({\bar P} - \Lambda)_{12}({\bm k_1}) ({\bar P} - \Lambda)_{23}({\bm k_2}) ({\bar P} - \Lambda)^+_{34}({\bm k_3}) ({\bar P} - \Lambda)^+_{41}({\bm k_4})\right]
\nonumber\\
&& \hskip 10pt +3\,\tr\left[({\bar P} - \Lambda)_{12}({\bm k_1}) ({\bar P} - \Lambda)^+_{23}({\bm k_2}) ({\bar P} - \Lambda)_{34}({\bm k_3}) ({\bar P} - \Lambda)^+_{41}({\bm k_4})\right]
\Bigr] \ ,
\label{eq:4.47c}
\eea
\bea
\Delta{\cal A}^{(2,2)} &=& \frac{-8}{V^3} \sum_{1,2,3,4} \sum_{{\bm k}_1,{\bm k}_2,{\bm k}_3,{\bm k}_4}
   \delta_{{\bm k}_1+{\bm k}_2+{\bm k}_3+{\bm k}_4,0}\,
    \varphi^{(4)}_{1234}({\bm k}_1,{\bm k}_2,{\bm k}_3)\,
\nonumber\\
&&\times \Bigl\{ \Phi^{-1}_{12}({\bm k}_1)\,\Phi^{-1}_{23}({\bm k}_2)\,
      \varphi^{-1}_{34}({\bm k}_3)\,\varphi^{-1}_{41}({\bm k}_4)\,    \tr\left[({\bar P} + {\bar P}^+ - \Lambda - \Lambda^+)_{12}({\bm k_1}) 
                      ({\bar P} + {\bar P}^+ - \Lambda - \Lambda^+)_{23}({\bm k_2}) \right.
\nonumber\\
&& \hskip 200pt \times \left.     \bigl(\qslash_{34}({\bm k}_3)\,\qslash^+_{41}({\bm k}_4) 
            + \qslash^+_{34}({\bm k}_3)\,\qslash_{41}({\bm k}_4)\bigr)\right]
\nonumber\\
&& \hskip 10pt + \Phi^{-1}_{12}({\bm k}_1)\,\varphi^{-1}_{23}({\bm k}_2)\,
      \Phi^{-1}_{34}({\bm k}_3)\,\varphi^{-1}_{41}({\bm k}_4)\,\tr\left[({\bar P} + {\bar P}^+ - \Lambda - \Lambda^+)_{12}({\bm k_1})\,
   \qslash_{23}({\bm k}_2) \right.
\nonumber\\
&& \hskip 200pt \times \left. ({\bar P} + {\bar P}^+ - \Lambda - \Lambda^+)_{34}({\bm k_3})\,
       \qslash^+_{41}({\bm k}_4)\right]
\Bigr\}\ ,
\label{eq:4.47d}
\eea
\bea
\Delta{\cal A}^{(0,4)} &=& \frac{-4}{V^3} \sum_{1,2,3,4} \sum_{{\bm k}_1,{\bm k}_2,{\bm k}_3,{\bm k}_4} \delta_{{\bm k}_1+{\bm k}_2+{\bm k}_3+{\bm k}_4,0}\,
   \varphi^{(4)}_{1234}({\bm k}_1,{\bm k}_2,{\bm k}_3)\,\varphi^{-1}_{12}({\bm k}_1)\,\varphi^{-1}_{23}({\bm k}_2)\,
   \varphi^{-1}_{34}({\bm k}_3)\,\varphi^{-1}_{41}({\bm k}_4)
\nonumber\\
&&\times \tr \left[ \qslash_{12}({\bm k_1})\,\qslash^+_{23}({\bm k}_2)\,\qslash_{34}({\bm k}_3)\,\qslash^+_{41}({\bm k}_4)\right]\ .
\label{eq:4.47e}
\eea
\ese
The relation between ${\bar P}$ and $P$ is
\be
{^0_r P}_{12}({\bm k}) = {^0_r{\bar P}}_{12}({\bm k}) - \Phi_{12}({\bm k})\,2\gamma T\sum_{3,4}\delta_{1-2,3-4}\,
   {^0_r q}_{34}({\bm  k})\ ,
\label{eq:4.48}
\ee
which replaces Eq.\ (\ref{eq:4.24a}). Note that the vertices in Eqs.\ (\ref{eqs:4.47}) are
invariant under cyclic permutations of the frequency and wave vector indices. Also note that these
vertices are in general not finite in the hydrodynamic limit of small wave vectors and frequencies, 
since the $n$-point
susceptibility $\varphi^{(n)}_{12\ldots n}({\bm k}_1,\ldots,{\bm k}_n)$ diverges in this 
limit.\cite{Belitz_Kirkpatrick_1997} This formulation therefore does not yield a local field
theory.

$P$ and $\Lambda$ can now be eliminated in analogy to Sec.\ \ref{subsubsec:IV.C.2}. 
The identities (\ref{eq:4.28})
turn into
\bse
\label{eqs:4.49}
\bea
(P - \Lambda)_{12}({\bm k}) &=&\Phi_{12}({\bm k})\, ({\hat\Gamma} P)_{12}({\bm k})
\nonumber\\
&=& \Phi_{12}({\bm k})\,({\hat\gamma}\Lambda)_{12}({\bm k})\ ,
\label{eq:4.49a}
\eea
with
\be
({\hat\Gamma} P)_{12}({\bm k}) = 2T\Gamma\sum_{3,4} \delta_{1-2,3-4}\,P_{34}({\bm k})\ .
\label{eq:4.49b}
\ee
\ese
$({\hat\gamma}\Lambda)$ is defined analogously, and so is $({\hat\gamma} q)$.
$\Lambda$ in terms of $q$ takes the form
\bea
\Lambda_{12}({\bm k}) &=& -2i\sum_3\frac{1}{V}\sum_{\bm p}\varphi^{(3)}_{132}({\bm p},{\bm k}-{\bm p})\,\varphi^{-1}_{13}({\bm p})\,\varphi^{-1}_{32}({\bm k}-{\bm p})
   \left[\qslash_{13}({\bm p})\,\qslash^+_{32}({\bm k}-{\bm p}) + \qslash^+_{13}({\bm p})\,\qslash_{32}({\bm k}-{\bm p})\right]
\nonumber\\
&& -2i\sum_3\frac{1}{V}\sum_{\bm p}\Phi^{(3)}_{132}({\bm p},{\bm k}-{\bm p}) \left[({\hat\gamma} q)_{13}({\bm p})\,({\hat\gamma} q^+)_{32}({\bm k}-{\bm p})
    + ({\hat\gamma} q^+)_{13}({\bm p})({\hat\gamma} q)_{32}({\bm k}-{\bm p}) \right.
\nonumber\\
&&\hskip 220pt \left. + {(\hat\gamma} q)_{13}({\bm p})({\hat\gamma} q)_{32}({\bm k}-{\bm p})\right]
+ O(q^3)\ ,
\label{eq:4.50}
\eea
which replaces Eq.\ (\ref{eq:4.29}). 

Note that here, and in Eqs.\ (\ref{eqs:4.51}) below, ${\hat\gamma} q$
always arises from a $\bar P$ that was shifted by means of Eq.\ (\ref{eq:4.48}), and therefore carries the
same implicit frequency restrictions as $P$ and $\Lambda$. That is, $({\hat\gamma} q)_{12}({\bm k})$ implies
$(n_1+1/2)(n_2+1/2) > 0$. A related remark is that the notation $\varphi^{(3)}_{123}$, $\varphi^{(4)}_{1234}$, 
etc. just specifies that
the convolution has hydrodynamic content, i.e., that all of the frequencies do not have the same sign, but not 
which frequencies are positive or negative. This information, which is necessary for calculating physical
quantities, can be obtained by an inspection of the matrix fields that multiply the convolution. 

The effective action analogous to Eqs.\ (\ref{eqs:4.32}) then reads
\bse
\label{eqs:4.51}
\be
{\cal A}_{\text{eff}} = {\cal A}_{\text{eff}}^{(2)} + \Delta{\cal A}_{\text{eff}}^{(3)}
+ \Delta{\cal A}_{\text{eff}}^{(4)} + O(q^5)\ .
\label{eq:4.51a}
\ee
with a Gaussian part
\be
{\cal A}_{\text{eff}}^{(2)} = -8\sum_{r=0,3} \sum_{1,2\atop 3,4} \frac{1}{V}\sum_{\bm k} {^0_r q}_{12}({\bm k})\,
     \left(\delta_{13}\delta_{24}\,\frac{1}{\varphi_{12}({\bm k})}
     - \delta_{1-2,3-4}\,2T\gamma\right)\,{^0_r q}_{34}(-{\bm k})
\label{eq:4.51b}
\ee
and nonlinearities
\bea
{\cal A}_{\text{eff}}^{(3)} &=& \frac{-4i}{V^{2}} \sum_{1,2\atop 3} 
   \sum_{{\bm k}_1,{\bm k}_2\atop {\bm k}_3} \delta_{{\bm k}_1+{\bm k}_2+{\bm k}_3,0} \,
\nonumber\\
&& \hskip 20pt \times    \Bigl\{ \varphi^{(3)}_{123}({\bm k}_1,{\bm k}_2)\,\varphi^{-1}_{23}({\bm k}_2)\,\varphi^{-1}_{31}({\bm k}_3) 
     \tr\left[({\hat\gamma} q + {\hat\gamma} q^+)_{12}({\bm k}_1)\,
                 \bigl(\qslash_{23}({\bm k}_2)\,\qslash^+_{31}({\bm k}_3)
              + \qslash^+_{23}({\bm k}_2)\,\qslash_{31}({\bm k}_3)\bigr)\right]
\nonumber\\
&& \hskip 35pt + \Phi^{(3)}_{123}({\bm k}_1,{\bm k}_2)\,  
       \tr\left[({\hat\gamma} q)_{12}({\bm k}_1)\,({\hat\gamma} q + {\hat\gamma} q^+)_{23}({\bm k}_2)\,
                       ({\hat\gamma} q^+)_{31}({\bm k}_3)\right]    \Bigr\}  \ ,
\label{eq:4.51c}
\eea
\bea
{\cal A}_{\text{eff}}^{(4)} &=& -\,\frac{4}{V^3} \sum_{1,2\atop 3,4} 
   \sum_{{\bm k}_1,{\bm k}_2\atop {\bm k}_3,{\bm k}_4}
     \delta_{{\bm k}_1+{\bm k}_2+{\bm k}_3 + {\bm k}_4,0}\,
      \varphi^{(4)}_{1234}({\bm k}_1,{\bm k}_2,{\bm k}_3)\,\varphi^{-1}_{12}({\bm k}_1)\,\varphi^{-1}_{23}({\bm k}_2)\,\varphi^{-1}_{34}({\bm k}_3)\,
                    \varphi^{-1}_{41}({\bm k}_4)\,
\nonumber\\
&& \hskip 200pt \times       \tr\left[\qslash_{12}({\bm k}_1)\,\qslash^+_{23}({\bm k}_2)\,
                      \qslash_{34}({\bm k}_3)\,\qslash^+_{41}({\bm k}_4)\right]
\nonumber\\
&&-\,\frac{4i}{V^{2}} \sum_{1,2\atop 3} 
   \sum_{{\bm k}_1,{\bm k}_2\atop {\bm k}_3} \delta_{{\bm k}_1+{\bm k}_2+{\bm k}_3,0}\,
      \varphi^{(3)}_{123}({\bm k}_1,{\bm k}_2)\, \varphi^{-1}_{23}({\bm k}_2)\, \varphi^{-1}_{31}({\bm k}_3)\,
\nonumber\\
&& \hskip 200pt \times    \tr\left[(\gamma\Lambda + \gamma\Lambda^+)_{12}({\bm k}_1)\,
         \bigl(\qslash_{23}({\bm k}_2)\,\qslash^+_{31}({\bm k}_3) + \qslash^+({\bm k}_2)\,
              \qslash_{31}({\bm k}_3)\bigr)\right]
\nonumber\\
&& -2 \sum_{12} \frac{1}{V}\sum_{\bm k} \tr\left[\Lambda_{12}({\bm k})\,{\hat\gamma}\Lambda^+_{21}({\bm k})
     \right]
\nonumber\\
&&-\,\frac{8}{V^3} \sum_{1,2\atop 3,4} \sum_{{\bm k}_1,{\bm k}_2\atop {\bm k}_3,{\bm k}_4} \delta_{{\bm k}_1+{\bm k}_2+{\bm k}_3 + {\bm k}_4,0}\,
      \varphi^{(4)}_{1234}({\bm k}_1,{\bm k}_2,{\bm k}_3)\, 
\nonumber\\
&& \hskip 50pt \times\Bigl\{
        \varphi^{-1}_{34}({\bm k}_3)\,\varphi^{-1}_{41}({\bm k}_4)\,
          \tr\left[({\hat\gamma} q + {\hat\gamma} q^+)_{12}({\bm k}_1)\,({\hat\gamma} q 
                            + {\hat\gamma} q^+)_{23}({\bm k}_2)\, \bigl(\qslash_{34}({\bm k}_3)\qslash^+_{41}({\bm k}_4) 
     + \qslash^+_{34}({\bm k}_3)\,\qslash_{41}({\bm k}_4)\bigr)\right]
\nonumber\\
&& \hskip 60pt + \varphi^{-1}_{23}({\bm k}_2)\,\varphi^{-1}_{41}({\bm k}_4)\,
     \tr\left[({\hat\gamma} q + {\hat\gamma} q^+)_{12}({\bm k}_1)\,\qslash_{23}({\bm k}_2)\,
         ({\hat\gamma} q + {\hat\gamma} q^+)_{34}({\bm k}_3)\,\qslash^+_{41}({\bm k}_4) \right] \Bigr\}
\nonumber\\
&&-\,\frac{4i}{V^{2}} \sum_{1,2\atop 3} 
   \sum_{{\bm k}_1,{\bm k}_2\atop {\bm k}_3} \delta_{{\bm k}_1+{\bm k}_2+{\bm k}_3,0}\,
      \Phi^{(3)}_{123}({\bm k}_1,{\bm k}_2,{\bm k}_3)\, \Bigl\{
        \tr\left[ ({\hat\gamma}\Lambda)_{12}({\bm k}_1)\,
              ({\hat\gamma} q + {\hat\gamma} q^+)_{23}({\bm k}_2)\,({\hat\gamma} q^+)_{31}({\bm k}_3)\right]
\nonumber\\
&&\hskip 192pt + \tr\left[ (\gamma q)_{12}({\bm k}_1)\,
        ({\hat\gamma}\Lambda + {\hat\gamma}\Lambda^+)_{23}({\bm k}_2)\,({\hat\gamma} q^+)_{31}({\bm k}_3)\right]
\nonumber\\
&& \hskip 192pt + \tr\left[ ({\hat\gamma} q)_{12}({\bm k}_1)\,
          ({\hat\gamma} q + {\hat\gamma} q^+)_{23}({\bm k}_2)\,({\hat\gamma}\Lambda^+)_{31}({\bm k}_3)\right]
          \Bigr\}
\nonumber\\
&&-\,\frac{2}{V^3} \sum_{1,2\atop 3,4} 
   \sum_{{\bm k}_1,{\bm k}_2\atop {\bm k}_3,{\bm k}_4} \delta_{{\bm k}_1+{\bm k}_2+{\bm k}_3 + {\bm k}_4,0}\,
       \Phi^{(4)}_{1234}({\bm k}_1,{\bm k}_2,{\bm k}_3)\, 
        \Bigl\{
        4\,\tr\left[ ({\hat\gamma} q)_{12}({\bm k}_1)\,({\hat\gamma} q)_{23}({\bm k}_2\,
                 ({\hat\gamma} q)_{34}({\bm k}_3)\,({\hat\gamma} q^+)_{41}({\bm k}_4)\right]
\nonumber\\
&& \hskip 145pt + 4\,\tr\left[ ({\hat\gamma} q^+)_{12}({\bm k}_1)\,({\hat\gamma} q^+)_{23}({\bm k}_2\,
          ({\hat\gamma} q^+)_{34}({\bm k}_3)\,({\hat\gamma} q)_{41}({\bm k}_4)\right]
\nonumber\\
&& \hskip 145pt + 3\,\tr\left[ ({\hat\gamma} q)_{12}({\bm k}_1)\,({\hat\gamma} q)_{23}({\bm k}_2\,
     ({\hat\gamma} q^+)_{34}({\bm k}_3)\,({\hat\gamma} q^+)_{41}({\bm k}_4)\right]
\nonumber\\
&& \hskip 145pt   + 3\,\tr\left[ ({\hat\gamma} q)_{12}({\bm k}_1)\,({\hat\gamma} q^+)_{23}({\bm k}_2\,
     ({\hat\gamma} q)_{34}({\bm k}_3)\,({\hat\gamma} q^+)_{41}({\bm k}_4)\right] \Bigr\}\ ,
\label{eq:4.51d}
\eea
\ese
where $\Lambda$ in terms of $q$ is given by Eq.\ (\ref{eq:4.50}).
The rules about treating $\qslash$ are the same as in the phase space formulation, and
the Gaussian contractions are
\bse
\label{eqs:4.52}
\bea
\langle{^0_r q}_{12}({\bm k})\,{^0_s q}_{34}(-{\bm k})\rangle &=& \langle{^0_r q}_{12}({\bm k})\,{^0_s \qslash}_{34}(-{\bm k})\rangle
   = \langle{^0_r \qslash}_{12}({\bm k})\,{^0_s q}_{34}(-{\bm k})\rangle
\nonumber\\
&=& \frac{V}{16}\,\delta_{rs}\left[\delta_{13}\,\delta_{24}\,\varphi_{12}({\bm k})
     + 2\gamma T\,\delta_{1-2,3-4}\,\frac{\varphi_{12}({\bm k})\,\varphi_{34}({\bm k})}
           {1 + 2\gamma\chi^{(0)}_{1-2}({\bm k})}\right]\ ,
\nonumber\\
\label{eq:4.52a}\\
\langle{^0_r \qslash}_{12}({\bm k})\,{^0_s \qslash}_{34}(-{\bm k})\rangle &=&
   \frac{V}{8}\,\gamma T\,\delta_{1-2,3-4}\,\frac{\varphi_{12}({\bm k})\,\varphi_{34}({\bm k})}
           {1 + 2\gamma\chi^{(0)}_{1-2}({\bm k})}\ ,
\label{eq:4.52b}
\eea
where
\be
\chi^{(0)}_{1-2}({\bm k}) = -T\sum_{34}\delta_{1-2,3-4}\,\varphi_{34}({\bm k})\ .
\label{eq:4.52c}
\ee
\ese
\end{widetext}

The above expressions provide all of the information necessary for one-loop renormalizations
of two-point correlation functions in the density formalism.

\section{A simple application: Perturbation theory for the density of states}
\label{sec:V}

As a simple demonstration of how the formalism developed above works, let us consider
the density of states (DOS). We first use the phase-space formalism from Sec.\ \ref{subsec:IV.D}. 
From Eq.\ (\ref{eq:2.17}) we see that the DOS is given by
\bse
\label{eqs:5.1}
\bea
N(\omega) &=& \NF + \Re Q(i\omega_n \to \omega + i0)
\nonumber\\
&\equiv& \NF + \delta N(\omega)\ ,
\label{eq:5.1a}
\eea
where $\NF$ is the DOS at the Fermi level including the interaction in Hartree-Fock
approximation, and
\be
Q(i\omega_n) = \frac{4}{\pi}\,\frac{1}{V} \sum_{\bm p} \left\langle {^0_0 P}_{nn}({\bm p},{\bm p})\right\rangle\ .
\label{eq:5.1b}
\ee
\ese
We next observe, from Eq.\ (\ref{eq:4.28}),
\be
\langle P_{11} \rangle = \langle\Lambda_{11}\rangle + \Phi_{11}\,\frac{2\gamma T}{V} 
     \sum_2 \langle\Lambda_{22}\rangle\ .
\label{eq:5.2}
\ee
The second contribution to $\langle P \rangle$ is proportional to the first one
with an additional frequency integration which decreases the effects of the soft modes. 
The leading hydrodynamic contribution to $Q$ is therefore given by Eq.\ (\ref{eq:5.1b})
with $P$ replaced by $\Lambda$. Using Eq.\ (\ref{eq:4.29}) we can express $Q$ in terms
of a loop expansion in terms of the $q$-correlation functions. To one-loop order the expression
for $\Lambda$ in Eq.\ (\ref{eq:4.29}) suffices. Of the two contributions to $\langle\Lambda_{11}\rangle$,
the second one has no hydrodynamic content due to the frequency restrictions inherent in
$\Phi$ or ${\hat\gamma} q$. From the first contribution we obtain
\be
Q(i\omega_{n_1}) = \frac{-2i}{\pi V} \sum_{{\bm p}_1} \sum_3 G_3^{-1}\,\tr \left\langle \qslash_{13}\,\qslash^+_{31} \right\rangle\ ,
\label{eq:5.3}
\ee
where we neglect 2-loop contributions and less divergent terms. Inserting the Gaussian propagator,
Eq.\ (\ref{eq:4.33b}), and choosing $n_1 > 0$, we find
\bea
Q(i\omega_{n_1}) &=& \frac{-2i\gamma}{\pi}\,\frac{1}{V} \sum_{\bm k} T\sum_{m<0} \varphi^{(3)\,++,-}_{n_1,n_1,m}({\bm 0},{\bm k})\,
\nonumber\\
&& \hskip 50pt \times    \frac{1}{1 + 2\gamma\chi^{(0)}_{n_1-m}({\bm k})}\ .
\label{eq:5.4}
\eea
Here $\varphi^{(3)\,++,-}$ is given by Eq.\ (\ref{eq:4.46b}), and we have explicitly indicated that the
first two frequency arguments are positive while the third one is negative, and $\chi^{(0)}$ is given by
Eq.\ (\ref{eq:4.52c}). The same result is obtained if we realize that $\delta N$ is given by 
$\langle P_{nn}({\bm p})\rangle$ in the density formalism of Sec.\ \ref{subsec:IV.E} and use
Eqs.\ (\ref{eq:4.49a}), (\ref{eq:4.50}), and (\ref{eq:4.52b}). 

To find the leading hydrodynamic contribution to $Q$ or $\delta N$ we can calculate $\varphi^{(3)}$ in 
the AGD approximation, whence it depends only on $n_1-m$. Performing the integral at $T=\omega=0$ in $d=1+\epsilon$ dimensions, we obtain
\bse
\label{eqs:5.5}
\be
\delta N(\omega=0)/\NF = \frac{-1}{\epsilon}\,\left[1 + O(\epsilon)\right]\,\frac{G}{\pi}\,f(A_0^{\text s}) + O(G^2)\ ,
\label{eq:5.5a}
\ee
where $G$ is the loop expansion parameter from Eq.\ (\ref{eq:4.36c}),
\be
f(x) = \frac{1-x/2}{\sqrt{1-x}} - 1\ ,
\label{eq:5.5b}
\ee
and
\be
A_0^{\text s} = 2\NF\gamma = \frac{2\NF\Gamma}{1 + 2\NF\Gamma}
\label{eq:5.5c}
\ee
\ese
is the  singlet s-wave Landau scattering amplitude in the particle-hole channel.

There are several interesting aspects of this result. The $1/\epsilon$ singularity
reflects the instability of the Fermi-liquid against a Luttinger liquid in $d=1$.\cite{Solyom_1979,
Castellani_DiCastro_Metzner_1994, Metzner_Castellani_DiCastro_1998, Giamarchi_2004} 
The current formalism emphasizes that this is a result of the
Goldstone fluctuations getting so strong that they destroy the ordered phase.
In fixed dimension $1<d<3$ as a function of the imaginary frequency the corresponding result
is
\bse
\label{eqs:5.6}
\be
(Q(i\omega_n) - Q(i0))/\NF = c_d\ f_d(A_0^{\text s})\,\omega_n^{d-1}\ ,
\label{eq:5.6a}
\ee
where $c_d>0$ interpolates smoothly between $c_d \propto 1/(d-1)$ for $d\to 1$ and
$c_d \propto 1/(3-d)$ for $d\to 3$. $f_d$ is a function that continuously evolves to $f$ given in
Eq.\ (\ref{eq:5.5b}) as $d\to 1$, is positive definite for $1<d\leq 3$, and has the property 
$f_d(x\to 0) \propto x^2$. For $d=3$
one finds
\bea
(Q(i\omega_n) - Q(i0))/\NF &=& {\tilde c}_3\ f_3(A_0^{\text s})\,\omega_n^{2}\log(1/\omega_n)
\nonumber\\
&&\hskip 40pt + O(\omega_n^2)
\label{eq:5.6b}
\eea
\ese
with ${\tilde c}_3>0$. 

After analytically continuing to the real axis these results imply for the leading nonanalytic
frequency correction to the density of states 
\bse
\label{eqs:5.7}
\be
\delta N(\omega)/\NF = n_d\,(2-d)\,\vert\omega\vert^{d-1} 
\label{eq:5.7a}
\ee
for $1<d<3$, and
\be
\delta N(\omega)/\NF = -{\tilde n}_3\,\omega^2 \log(1/\vert\omega\vert) 
\label{eq:5.7b}
\ee
\ese
for $d=3$. $n_d$ is positive definite and interpolates smoothly between $n_d \propto 1/(d-1)$ 
for $d\to 1$ and $n_d \propto 1/(3-d)$ for $d\to 3$, and ${\tilde n}_3>0$.
 
Note that to one-loop order there is no $\vert\omega\vert$ contribution to the DOS in 
$d=2$, in agreement with earlier results obtained with diagrammatic
techniques.\cite{Mishchenko_Andreev_2002, Chubukov_et_al_2005, 
Chubukov_Maslov_2012, long_range_footnote}
Equations (\ref{eq:5.5a}), (\ref{eqs:5.6}), and (\ref{eqs:5.7}) are analogous to the Coulomb or zero-bias 
anomaly in a disordered Fermi liquid.\cite{Altshuler_Aronov_1979,
Lee_Ramakrishnan_1985, Belitz_Kirkpatrick_1994} There, the singularity at zero
frequency is proportional to $1/(d-2)$, and the nonanalytic frequency 
dependence is $\vert\omega\vert^{(d-2)/2}$.
The exponents of the respective frequency nonanalyticities reflect the scale dimensions
of the least irrelevant operator at the disordered and clean Fermi-liquid fixed point,
which are $-(d-2)$ and $-(d-1)$, and the corresponding dynamical exponents, which
are $z=2$ and $z=1$, respectively.\cite{Castellani_DiCastro_Metzner_1994, 
Belitz_Kirkpatrick_1997, Metzner_Castellani_DiCastro_1998} If we denote the
least irrelevant operator by $u$, this leads to a homogeneity law in the clean case
\be
\delta N(\omega) = \delta N(\omega\,b, u\,b^{-(d-1)})\ ,
\label{eq:5.8}
\ee
which immediately leads to the frequency dependence reflected in Eq.\ (\ref{eq:5.7a}).
The sign of the effect at zero frequency, Eq.\ (\ref{eq:5.5a}), 
reflects the fact that a repulsive interaction will lead to a decrease of the density
of states at the Fermi level.

The dependence of the anomaly on the interaction amplitude also warrants a comment.
The function $f(x)$, Eq.\ (\ref{eq:5.5b}), is positive definite for $0<x<1$, and 
$f(x\to 0) \propto x^2$. The effects of the soft modes thus appear only at second
order in the interaction, whereas Eq.\ (\ref{eq:5.3}) or (\ref{eq:5.4}) naively suggest
an effect at first order. The technical reason is that the integral in Eq.\ (\ref{eq:5.4})
vanishes if one puts $\gamma = 0$ in the denominator of the integrand. This is
consistent with many-body perturbation theory, as it must be:
\begin{figure}[t]
\vskip -0mm
\includegraphics[width=8.0cm]{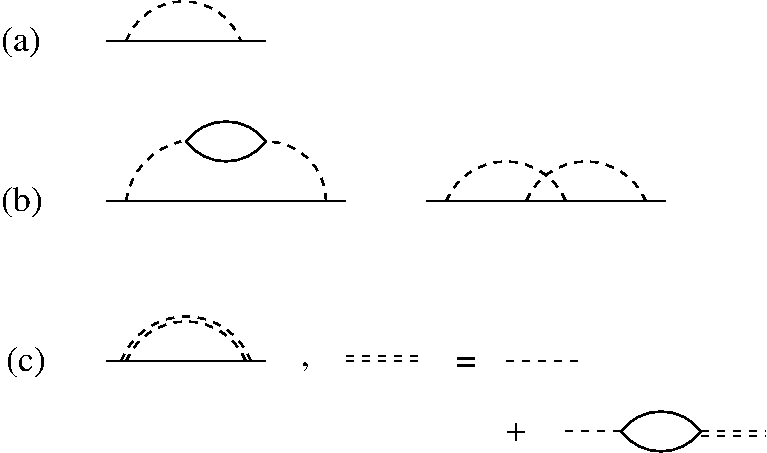}
\caption{Diagrammatic contributions to the DOS at (a) first and (b) second order
               in the interaction amplitude. (c) Shows a ladder resummation of the
               diagram in (a) plus the first diagram in (b), which is identical with Eq.\ (\ref{eq:5.4}).}
\label{fig:5.1}
\end{figure}
The diagram in Fig.\ \ref{fig:5.1} (a), which is the only contribution to the DOS to first
order in the interaction (Hartree diagrams have been absorbed into the Green function),
does not allow for a mixing of retarded and advanced degrees of freedom and hence
has no hydrodynamic content. The first diagram in Fig.\ \ref{fig:5.1} (b) does, and its
hydrodynamic part is identical with Eq.\ (\ref{eq:5.4}) to second order in $\gamma$.
The hydrodynamic part of its ladder resummation, Fig.\ \ref{fig:5.1} (c), is identical with
the full Eq.\ (\ref{eq:5.4}). The other contribution to second order in the interaction,
i.e., the second diagram in Fig.\ \ref{fig:5.1} (b), represents a particle-particle channel
contribution. In the present formalism it will appear if a particle-particle interaction
amplitude is added to Eqs.\ (\ref{eq:2.4b}) or (\ref{eqs:2.13}).\cite{us_tbp} Note that the
effective theory picks out only the hydrodynamic parts of these diagrams, whereas
the many-body diagrams contain non-hydrodynamic pieces as well. Also note that
the current theory, at any given order in the loop expansion, in valid to all orders
in the interaction amplitude, whereas many-body perturbation theory by necessity
involves an expansion in powers of the interaction.

In addition to the above discussion there are other interesting aspects
of the DOS that include the behavior in the case of a long-ranged Coulomb 
interaction,\cite{Khveshchenko_Reizer_1998, Mishchenko_Andreev_2002}
as well as the two-point correlation function of the local DOS. These effects will be
discussed elsewhere.\cite{us_tbp}

\section{Summary, Discussion, and Outlook}
\label{sec:VI}

\subsection{Summary and Discussion}
\label{subsec:V.A}

We have derived an effective field theory that describes the soft modes, and their effects
on observables, in electron systems without quenched disorder. The soft modes have
been identified by means of a Ward idendity; they are particle-hole excitations with a
linear frequency-momentum relation. They are the Goldstone modes of a continuous
symmetry, namely, a rotation in Matsubara frequency space that relates retarded and
advanced degrees of freedom and is spontaneously broken whenever the quasiparticle
spectral weight is nonzero. This identifies a Fermi liquid as an ordered phase with the
quasiparticle spectral weight as the order parameter and the soft particle-hole excitations
the corresponding Goldstone modes. A nonzero temperature the Goldstone modes
acquire a mass, and so do subsets of them in the presence of a magnetic field or
other symmetry-breaking external fields. Quenched disorder restricts the Goldstone
modes to the s-wave or $\ell=0$ channel, whereas all higher angular momentum 
channels become massive. This is the reason why the clean theory is more complicated
than the corresponding one for disordered electrons: There are many more soft modes
to take into account in clean systems. It is important to remember that the Goldstone 
modes are {\em not} density excitations, and their existence is not related to particle-number
conservation; the density susceptibility is characterized by zero-sound excitations
regardless of whether the symmetry is broken or not, and at any temperature.

Technically, we have borrowed heavily from techniques and concepts developed by
Wegner\cite{Wegner_1979} for disordered noninteracting electrons and generalized by 
Finkelstein\cite{Finkelstein_1983} to the interacting disordered case. However, the
resulting effective field theory in the clean case does {\em not} take the form of a
generalized nonlinear sigma model, as it does in the disordered case, but is significantly more
complex. This complexity reflects the presence of many more soft modes. One 
manifestation of this complexity is the fact that the effective action contains terms
to higher than linear order in the interaction amplitude, whereas in the disordered
case only linear terms appear. The reason is that in the clean case the Green function
is soft, and the additional powers of the frequency that accompany higher powers of
the interaction are compensated by multiplicative Green functions. In the disordered
case, in contrast, the Green function is massive because of the elastic relaxation rate,
and terms of higher than linear order in the interaction are irrelevant in a
renormalization-group sense. 

We have followed Wegner\cite{Wegner_1979} and later work based on his
ideas\cite{Schaefer_Wegner_1980, Efetov_Larkin_Khmelnitskii_1980, 
Pruisken_Schaefer_1982, Belitz_Kirkpatrick_1994, Belitz_Kirkpatrick_1997} in 
bosonizing the theory by
introducing classical matrix fields that are isomorphic to bilinear products of
fermion fields. Technically, this is done by introducing a Lagrange multiplier
constraint and integrating out the fermions. This approach is very different
from that of Shankar,\cite{Shankar_1994} who applied renormalization-group
techniques directly to the fermionic theory to derive Landau Fermi-liquid theory.
The Lagrange multiplier field, whose expectation value plays the role of a
self energy, is soft and needs to be kept to all orders to avoid introducing
spurious soft modes. This has been achieved in terms of diagram rules for
the effective theory, with the role of the Lagrange multiplier being to eliminate
the noninteracting part of certain propagators. The effective action has not
been given in closed form, but an explicit method has been devised to 
construct it to any desired power in the fundamental soft-mode field. This
allows for a systematic loop expansion, with the inverse Fermi velocity
playing the role of the bare loop expansion parameter. The interaction, by
contrast, is {\em not} treated perturbatively. Order by order in a loop expansion,
the theory thus produces the leading hydrodynamic (i.e., long-wavelength and
low-frequency) effects to all orders in the interaction, and the interaction is
not required to be weak in any sense. This, together with its ability to isolate
the hydrodynamic contributions to any desired observable, sets the current
theory apart from many-body perturbation theory which expands in powers
of the interaction, with any resummations to all orders in the interaction
being opportunistic and not controlled.

A related feature of the current theory that is even more important is its
amenability to renormalization-group techniques. Since it is formulated
in terms of an infinite power series in the fundamental field, with each term
characterized by an n-point vertex and a corresponding correlation function,
the structure of which is dictated by symmetry, the bare theory also suggests the
structure of the corresponding renormalized theory, which many-body
perturbation theory cannot do. This feature has been exploited extensively
in the disordered case to describe, inter alia, the Anderson-Mott metal-insulator
transition,\cite{Finkelstein_1984a, Finkelstein_1984b, Belitz_Kirkpatrick_1994}
which had proven very hard to do by means of many-body theory. 

Certain aspects of the present theory had already been present in the theory
put forward in Ref.\ \onlinecite{Belitz_Kirkpatrick_1997}. However, since the 
latter focused on disordered systems, it was formulated in terms of
s-wave variables, with the higher angular momentum modes effectively
being integrated out. In an application to the clean case, this led to a
nonlocal theory, i.e., the 3-point and higher vertices did not exist in the
hydrodynamic limit of small frequencies and wave numbers. In addition,
no complete separation of soft and massive modes was achieved in that
theory for the clean case. Both of these points have been remedied in the
present formulation. The issue of the higher-angular-momentum modes
is handled by formulating the theory in terms of phase-space variables,
which explicitly keeps all angular momentum channels, and the separation
of soft and massive modes has been achieved by expressing the massive
modes in terms of the soft ones in a power-series expansion, albeit not
in a closed form, in contrast to the disordered case. The relation to the
formulation in terms of s-wave variables, which is still useful for certain
calculational purposes, has been explained in Sec.\ \ref{subsec:IV.E}.

\subsection{Outlook}
\label{subsec:V.B}

Formalism for formalism's sake is not very useful, and a crucial test for
the present theory will be any physical problems it can solve more easily
than other methods. We conclude this paper by listing several applications
that we plan to pursue in the near future, many of which we already have
obtained preliminary results for.

The density-of-states calculation in Sec.\ \ref{sec:V}
is meant only as an illustration of how the loop-expansion-based
perturbation theory works within the present context; it does not achieve
anything that cannot easily be done with many-body diagrams. It
nevertheless has some important advantages, and points to some
crucial future developments. For instance, as was shown in 
Ref.\ \onlinecite{Belitz_Kirkpatrick_1997},
the Fermi-liquid state is represented by a stable renormalization-group fixed
point, and the leading irrelevant operators with respect to that fixed point
can be easily identified. This shows that the nonanalytic energy or frequency
dependence of the DOS near the Fermi level, Eqs.\ (\ref{eqs:5.6}), is qualitatively
exact; there cannot be any contributions with a lower power of the frequency.
This would be impossible to establish within many-body theory. A related
point is that the field theory, in contrast to many-body perturbation theory,
is not perturbative in the interaction; the results at any order in the loop
expansion are valid to all orders in the interaction. Other observables
will also display universal nonanalytic behavior in the Fermi-liquid phase that
can be understood as corrections to scaling at the Fermi-liquid fixed point,
and they can be identified by the same techniques. One interesting quantity
is the two-point correlation function of the local DOS. Since the DOS is the
(zeroth moment of) the order parameter for the Fermi liquid, this plays the
role of the order parameter susceptibility, and it is expected to be long-ranged,
i.e., to exhibit a universal divergence for small wave numbers, everywhere in
the Fermi-liquid phase. This expected phenomenon is analogous to a
well-known feature of the magnetic susceptibility in the ordered phase of a 
Heisenberg ferromagnet.\cite{Brezin_Wallace_1973} In Ref.\ \onlinecite{Kirkpatrick_Belitz_2011b}
we have used scaling arguments combined with a preliminary version of the
current theory to predict that the DOS-DOS correlation function diverges as
$T/k^{3-d}$ for small wave numbers $k$. We will revisit this problem, together
with other aspects of the DOS, in more detail shortly.

In addition, the effective theory suggests the existance of a critical fixed point
that describes a transition from a Fermi liquid to a non-Fermi-liquid state.
This will be analogous in many respects to the critical fixed points that
describe various versions of the Anderson-Mott transition.\cite{Belitz_Kirkpatrick_1994}
The basic physical idea is that the fluctuations in a Fermi liquid that are
represented by the Goldstone modes become stronger with decreasing
dimensionality and increasing interaction strength, and ultimately give rise
to an instability of the Fermi liquid since the system can lower its free energy
by going into a state that restores the symmetry and thus eliminates the
Goldstone modes. A scaling theory for such a transition has been developed
in Ref.\ \onlinecite{Kirkpatrick_Belitz_2011b}, and arguments have been given
that suggest that the corresponding fixed point will appear in the present theory
at 2-loop order. The loop expansion will then allow for a controlled description
of the transition in $d=1+\epsilon$ dimensions.

The theory can also be applied to magnetic phenomena, i.e., order in the
spin-triplet channel. In Sec.\ \ref{subsubsec:IV.A.2} we have constructed a
saddle-point solution that reproduces the Stoner theory of ferromagnetism.
By expanding about that saddle point, analogously to the expansion about the
Fermi-liquid saddle point in Sec.\ \ref{subsec:IV.B}, one can construct a
systematic theory for quantum Heisenberg ferromagnets that improves upon
and replaces the Hertz-Millis theory.\cite{Hertz_1976, Millis_1993} Such
a theory has been developed in Ref.\ \onlinecite{Kirkpatrick_Belitz_2003b},
and it has been shown to have very interesting consequences, including the
fact that the phase transition at sufficiently low temperatures in a Heisenberg
ferromagnet is generically of first order, in agreement with experimental
observations. This theory was based on a phenomenological treatment of
the fermionic degrees of freedom combined with symmetry arguments, and it
does not have the correct limiting behavior as $d\to 1$.
The current theory will allow for a systematic derivation valid in any dimension
$d>1$ that will serve as an important check and will probably point to new
developments. It can also be applied to ferrimagnets, the quantum aspects of
which have not been considered so far. Finally, if one considers band electrons
instead of the nearly-free electron model we have employed for simplicity, the
theory can also describe antiferromagnetism.

Treatments of the spin-triplet channel are not restricted to s-wave order. By
keeping a triplet interaction amplitude in the p-wave channel, see Eq.\ (\ref{eq:2.13d}),
one can describe a magnetic nematic or p-wave ferromagnet. In
Ref.\ \onlinecite{Kirkpatrick_Belitz_2011a} we have shown that the quantum phase
transition from a Fermi liquid to such magnetic state is generically of first order,
for reasons that are similar to those for the analogous statement for the s-wave
ferromagnetic transition. This is in contrast to Hertz-type 
theories\cite{Oganesyan_Kivelson_Fradkin_2001} that predict a continuous transition.
This result was also based on a preliminary version of the current theory, and the
present complete formulation will allow for more detailed studies of exotic
magnetic states. The same is true for superconducting states, in either the
s-wave or higher angular momentum channels.

All of the above examples involve phenomena that crucially rely on the coupling
of some order parameter to fermionic soft modes. The present effective theory
provides a general framework for describing the universal aspects of ordered
phases, and the quantum phase transitions between them, of clean electron systems.

\acknowledgments
We gratefully acknowledge helpful correspondence with Andrey Chubukov and 
Dmitrii Maslov regarding the density of states in 2-d systems.
This work was supported by the National Science Foundation under Grant Nos.
DMR-09-29966, and DMR-09-01907.

\appendix

\section{Effects of quenched disorder}
\label{app:A}

In this appendix we show that quenched disorder modifies the Ward
identity we discussed in Sec.\ \ref{sec:III} such that only density or
$\ell=0$ modes remain soft. 

Consider a static random potential $u({\bm x})$ that couples to the
electron number density and has a Gaussian distribution with a second
moment
\be
\left\{u({\bm x})\,u({\bm y})\right\}_{\text{dis}} = U({\bm x}-{\bm y})/\pi\NF\ ,
\label{eq:A.1}
\ee
where $\left\{ \ldots \right\}_{\text{dis}}$ denotes the disorder average and $U$ is
dimensionally an energy density. We then need to add to Eq.\ (\ref{eq:2.13a}) a 
contribution
\bea
{\cal A}_{\text{dis}} &=& \frac{-1}{2\pi\NF} {\sum_{\bm q}}^{\prime} U({\bm q}) \sum_{\bm k}
   \tr \left(Q({\bm k};{\bm q})\right)\, \tr\left(Q({\bm p};-{\bm q})\right)
\nonumber\\
&& + \frac{1}{\pi\NF} {\sum_{\bm q}}^{\prime} \sum_{{\bm k},{\bm p}} U({\bm k}-{\bm p})\,
   \tr\left(Q({\bm k};{\bm q})\,Q({\bm p};-{\bm q})\right)\ .
\nonumber\\
\label{eq:A.2}
\eea
The trace now includes tracing over replica indices to handle the disorder average.
Now we anticipate that the center-of-mass momenta ${\bm k}$ and ${\bm p}$ will
be pinned to the Fermi surface, and expand $U({\bm k}-{\bm p})$ in Legendre
polynomials. We then can write the disorder part of the action
\bse
\label{eqs:A.3}
\bea
{\cal A}_{\text{dis}} &=& \frac{-1}{2\pi\NF\tau_1} {\sum_{\bm q}}^{\prime} 
   \tr\left(Q^{0,0}({\bm q})\right)\,\tr\left(Q^{0,0}(-{\bm q})\right)
\nonumber\\
&& + \frac{1}{\pi\NF} \sum_{\ell=0}^{\infty} \frac{1}{\tau_{\text{rel}}^{(\ell)}}\,
   \frac{1}{\kF^{2\ell}} \sum_{m=-\ell}^{\ell} (-)^m {\sum_{\bm q}}^{\prime} 
\nonumber\\
&& \hskip 30pt \times\tr\left(
   Q^{\ell,m}({\bm q})\,Q^{\ell,-m}(-{\bm q})\right)\ .
\label{eq:A.3a}
\eea
Here we have expanded the $Q$-matrices in spherical harmonics $Y_{\ell}^m$,
\be
Q_{12}^{\ell,m}({\bm q}) = \sum_{\bm k} k^{\ell}\,
   Y_{\ell}^{m}(\Omega_{\bm k})\,Q_{12}({\bm k};{\bm q})\ ,
\label{eq:A.3b}
\ee
and have defined relaxation rates $1/\tau_1 = U({\bm q}=0)$ and
\be
\frac{1}{\tau_{\text{rel}}^{(\ell)}} = 2\pi \int_{-1}^{1} d\eta\,P_{\ell}(\eta)\,U(\kF\sqrt{2(1-\eta)})
\label{eq:A.3c}
\ee
in the angular momentum channels.
\ese

The derivation of the Ward identity as in Sec.\ \ref{sec:III} now still yields Eq.\ (\ref{eq:3.12})
(note $Q^{0,0} \equiv Q^{(0)}$), but it gives no information about the angular momentum
channels with $\ell > 1$. This is consistent with the results of perturbation theory. The
Fermi-liquid saddle-point solution from Sec.\ \ref{subsubsec:IV.A.1} needs to be
augmented by a disorder-induced self energy and becomes identical with the 
saddle point considered in Ref.\ \onlinecite{Belitz_Kirkpatrick_1997}. An expansion about
the saddle point to Gaussian order yields
\bea
\langle\delta Q_{12}^{\ell,m}({\bm q}=0)\,Q_{12}^{\ell,m}({\bm q}=0)\rangle_{i\Omega_{n_1-n_2}\to 0} & &
\nonumber\\
&& \hskip -90pt\propto
   \frac{\pi\NF}{1/\tau_{\text{rel}}^{(\ell=0)} - 1/\tau_{\text{rel}}^{(\ell)}}\ .
\label{eq:A.4}
\eea
In order for the theory to be stable we need to require $1/\tau_{\text{rel}}^{(\ell=0)} > 1/\tau_{\text{rel}}^{(\ell>0)}$,
which will generically be the case. If it is not one needs to choose a saddle point in 
the angular-momentum channel that has the largest value of $1/\tau_{\text{rel}}^{(\ell=0)}$.
We see that all correlation functions other than the one for $\ell = 0$ are massive, which is
consistent with the information obtained from the Ward identity.

\section{An illustrative example: Soft modes in the classical
$O(2)$ vector model}
\label{app:B}

In this appendix we discuss a classical $O(2)$ vector model as an illustrative example of how
to separate soft and massive modes. Consider a 2-component classical field
${\bm\phi}({\bm x}) = (\pi({\bm x}),\sigma({\bm x}))$ and an action
\be
{\cal A} = \int d{\bm x}\ \left[ \frac{r}{2}\,{\bm\phi}^2({\bm x}) + \frac{c}{2}\,\left(\nabla{\bm\phi}({\bm x})\right)^2
   + \frac{u}{4}\,\left({\bm\phi}^2({\bm x})\right)^2 \right]
\label{eq:B.1}
\ee
with $(\nabla{\bm\phi})^2 = \partial_i \phi_j \partial^i \phi^j$. The partition function is
\be
Z = \int D[{\bm\phi}]\ e^{-{\cal A}[{\bm\phi}]}\ .
\label{eq:B.2}
\ee
The saddle-point equations for homogeneous field configurations $(\pi,\sigma)$ read
\bse
\label{eqs:B.3}
\bea
r\pi + u\pi^3 + u\pi\sigma^2 &=& 0\ ,
\label{eq:B.3a}\\
r\sigma + u\sigma^3 + u\sigma\pi^2 &=& 0\ .
\label{eq:B.3b}
\eea
\ese
For $r<0$, a solution that minimizes the free energy is
\bse
\label{eqs:B.4}
\bea
\pi_{\text{sp}} &=& 0\ ,
\label{eq:B.4a}\\
\sigma_{\text{sp}} &=& \sqrt{-r/u} \equiv \phi_0\ .
\label{eq:B.4b}
\eea
\ese
Alternatively, we can choose $\pi_{\text{sp}}$ to have any value with $\vert\pi_{\text{sp}}\vert < \phi_0$.
Then
\be
\sigma_{\text{sp}}^2 = \phi_0^2 (1-\pi_{\text{sp}}^2/\phi_0^2)
\label{eq:B.5}
\ee
also solves the saddle-point equations. This reflects the $O(2)$ rotational symmetry of the model.

\begin{figure}[t]
\vskip -0mm
\includegraphics[width=8.0cm]{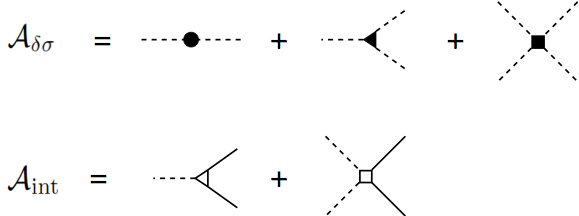}
\caption{Graphic representation of the contributions ${\cal A}_{\delta\sigma}$ and
               ${\cal A}_{\text{int}}$ to the action. Dashed and solid lines represent the
              fields $\delta\sigma$ and $\pi$, respectively.}
\label{fig:A_1}
\end{figure}

Now consider fluctuations about the saddle point, $\sigma = \phi_0(1 + \delta\sigma)$, and redefine
$\pi \to \phi_0\pi$, so that $\delta\sigma$ and $\pi$ are both dimensionless. Then the action takes 
the form
\bse
\label{eqs:B.6}
\be
{\cal A} = {\cal A}_{\text{sp}} + {\cal A}_{\pi}[\pi] + {\cal A}_{\delta\sigma}[\delta\sigma] 
                 + {\cal A}_{\text{int}}[\pi,\delta\sigma]\ ,
\label{eq:B.6a}
\ee
where
\bea
{\cal A}_{\pi}[\pi] &=& \frac{c}{2}\,\phi_0^2 \int d{\bm x}\ \left({\bm\nabla}\pi\right)^2
   + \frac{u}{4}\,\phi_0^4 \int d{\bm x}\ \pi^4\ ,
\nonumber\\
\label{eq:B.6b}\\
{\cal A}_{\delta\sigma}[\delta\sigma] &=& u\phi_0^4 \int d{\bm x}\ \left(\delta\sigma\right)^2
   + \frac{c}{2}\,\phi_0^2 \int d{\bm x}\ \left({\bm\nabla}\delta\sigma\right)^2 
\nonumber\\
&&   + u\phi_0^4 \int d{\bm x}\ (\delta\sigma)^3 + \frac{u}{4}\,\phi_0^4 \int d{\bm x}\ (\delta\sigma)^4\ ,
\nonumber\\
\label{eq:B.6c}\\
{\cal A}_{\text{int}}[\pi,\delta\sigma] &=& u\phi_0^4 \int d{\bm x}\ \pi^2\,\delta\sigma
   + \frac{u}{2}\,\phi_0^4 \int d{\bm x}\ \pi^2 (\delta\sigma)^2\ .
\nonumber\\
\label{eq:B.6d}
\eea
\ese
Here $\delta\sigma$ and $\pi$ are functions of ${\bm x}$. A graphic representation of the terms that
depend on $\delta\sigma$ is shown in Fig.\ \ref{fig:A_1}.

It is well known that the transverse fluctuation, i.e. the field $\pi$, is the soft mode in this 
model.\cite{Zinn-Justin_1996} $\pi$ is the Goldstone mode that reflects the spontaneously
broken $O(2)$ invariance, and it is governed by a Ward identity that ensures that the
homogeneous transverse susceptibility diverges. The effective soft-mode theory is a
nonlinear sigma model with an action
\bse
\label{eqs:B.7}
\be
{\cal A}_{\text{NL$\sigma$M}} = \frac{c}{2}\,\phi_0^2 \int d{\bm x}\ \left[({\bm\nabla}\pi)^2 
   +({\bm\nabla}\sqrt{1-\pi^2})^2\right]
\label{eq:B.7a}
\ee
and a partition function
\be
Z = \int \frac{D[\pi]}{\sqrt{1-\pi^2}}\,e^{-{\cal A}_{\text{NL$\sigma$M}}[\pi]}\ .
\label{eq:B.7b}
\ee
\ese
Notice that the functional integration measure has changed by switching to $\pi$ as the
only field. This is important to preserve the symmetry of the problem. Technically, terms
arising from the measure cancel spurious mass terms that get generated in perturbation
theory.

In this simple case the nonlinear sigma model can be derived in closed form by
writing ${\bm\phi}$ as a representation of $O(2)$ rotations,
\be
{\bm\phi}({\bm x}) = \left(\begin{array}{cc} \sqrt{1-\pi^2({\bm x})} & \pi({\bm x}) \cr
                                                                    -\pi({\bm x}) & \sqrt{1-\pi^2({\bm x})} 
                                         \end{array}\right)\,
                                  \left(\begin{array}{c} 0 \cr
                                                                   \rho({\bm x})
                                          \end{array}\right) \ ,
\label{eq:B.8}
\ee
and integrating out the massive $\rho$-fluctuations in saddle-point approximation
(i.e., replacing $\rho({\bm x})$ by its saddle-point value $\phi_0$,
see Ref.\ \onlinecite{Zinn-Justin_1996}). For the more complicated matrix field theory we
are interested in no closed-form derivation of the effective soft-mode theory has been
found so far, but one can still derive the latter perturbatively order by order in powers
of the soft modes. To guide that derivation it is useful to consider the analogous process
for the $O(2)$ model. In what follows, we thus pretend that the nonlinear sigma model
were not known, and construct an effective soft-mode theory perturbatively by 
considering Eqs.\ (\ref{eqs:B.6}) and systematically integrating out $\delta\sigma$.

Returning to Eqs.\ (\ref{eqs:B.6}) we see that at the Gaussian level $\pi$ is indeed massless and 
$\delta\sigma$ is massive. However, the $\pi^4$ term in Eq.\ (\ref{eq:B.6b}) does not carry
any gradients. If we were to simply neglect $\delta\sigma$ this term would generate a
mass for $\pi$ at the one-loop level. This mass violates the Ward identity and is spurious;
it gets canceled if one properly integrates out $\delta\sigma$. The coupling between
$\delta\sigma$ and $\pi$ is thus crucial and needs to be kept. Formally integrating out
$\delta\sigma$ leads to an effective action in terms of $\pi$ only,
\bse
\label{eqs:B.9}
\be
{\cal A}_{\text{eff}}[\pi] = {\cal A}_{\pi}[\pi] + \Delta{\cal A}_{\pi}[\pi]
\label{eq:B.9a}
\ee
where
\be
\Delta{\cal A}_{\pi}[\pi] = -\ln \int D[\delta\sigma]\ e^{-{\cal A}_{\delta\sigma}[\delta\sigma]
   - {\cal A}_{\text{int}}[\pi,\delta\sigma]}\ .
\label{eq:B.9b}
\ee
\ese
We can now perform the $\delta\sigma$-integral perturbatively. 

\begin{figure}[t]
\vskip -0mm
\includegraphics[width=8.0cm]{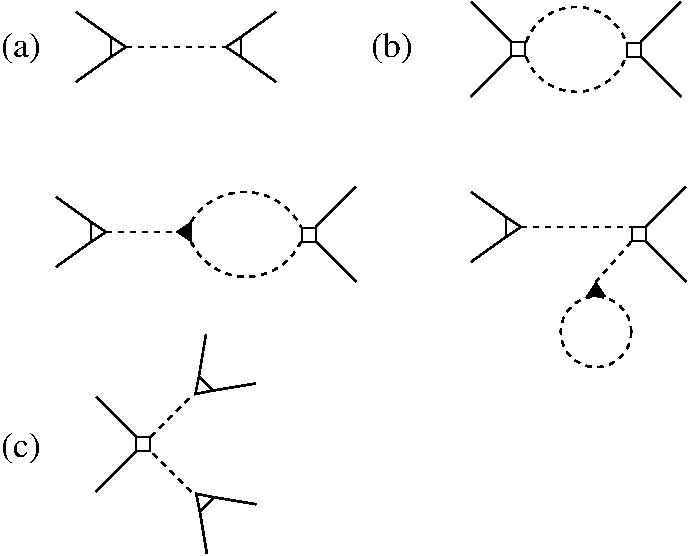}
\caption{(a) Tree-level and (b) one-loop contributions to the term of $O(\pi^4)$ in
              ${\Delta\cal A}_{\pi}[\pi]$ in an expansion in $\delta\sigma$ loops. (c) Tree-level
               contribution to the term of $O(\pi^6)$.}
\label{fig:A_2}
\end{figure}

Let us first integrate out $\delta\sigma$ in a straightforward manner. The resulting loop expansion 
is an expansion in powers of $1/\phi_0$. The tree-level and one-loop diagrams that contribute
to $O(\pi^4)$ are shown
in Fig.\ \ref{fig:A_2}. 
The tree diagram yields
\be
\Delta{\cal A}_{\pi}[\pi] \approx -\frac{u}{4}\,\phi_0^4 \int d{\bm x}\ \pi^4 + 
   \frac{c}{8}\,\phi_0^2 \int d{\bm x}\ ({\bm\nabla}\pi^2)^2
\label{eq:B.10}
\ee
Combining Eq.\ (\ref{eq:B.10}) with Eq.\ (\ref{eq:B.6b}) we see that the offending
$\pi^4$ term cancels, and the gradient terms agree with the expansion of the
nonlinear sigma model, Eq.\ (\ref{eq:B.7a}), to order $\pi^4$. Terms of higher
order in $\pi$ can be constructed analogously, see Fig.\ \ref{fig:A_2}(c).

\begin{figure}[t]
\vskip -0mm
\includegraphics[width=8.0cm]{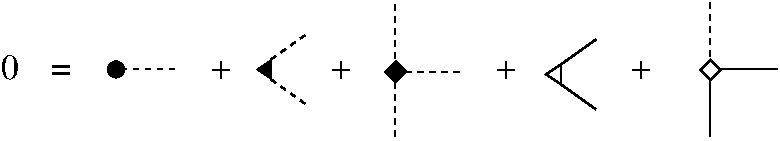}
\caption{Saddle-point equation for $\delta\sigma$ for fixed $\pi$.}
\label{fig:A_3}
\end{figure}

Alternatively, we can evaluate the integral in Eq.\ (\ref{eq:B.9b}) by means of a saddle-point
approximation for $\delta\sigma$ for fixed $\pi$. The saddle-point equation is shown
graphically in Fig.\ \ref{fig:A_3}. 
Analytically, we obtain
\bse
\label{eqs:B.11}
\bea
0 &=& \frac{1}{2}\,\pi^2({\bm x}) 
       + \left(1 + \frac{1}{2}\,\pi^2({\bm x})\right) \delta\sigma_{\text{sp}}({\bm x}) 
\nonumber\\
&&   - \frac{c}{2u\phi_0^2}\,{\bm\nabla}^2 \delta\sigma_{\text{sp}}({\bm x})
       + \frac{3}{2}\,\left(\delta\sigma_{\text{sp}}({\bm x})\right)^2
      + \frac{1}{2}\left(\delta\sigma_{\text{sp}}({\bm x})\right)^3.
\nonumber\\
\label{eq:B.11a}
\eea
In a gradient expansion, and keeping terms up to $O(\nabla^2)$, this is solved by
\bea
\delta\sigma_{\text{sp}}({\bm x}) &=& \sqrt{1-\pi^2({\bm x})} - 1 
\nonumber\\
&& + \frac{g}{1-\pi^2({\bm x})}\,{\bm\nabla}^2 \sqrt{1-\pi^2({\bm x})} + O(g^2{\bm\nabla}^4)\ ,
\nonumber\\
\label{eq:B.11b}
\eea
\ese
with $g = c/2u\phi_0^2$. Note that higher powers of $g$ are accompanied by higher
powers of gradients. Substituting this result back into the action, Eqs.\ (\ref{eqs:B.6}), 
we obtain the action of the nonlinear sigma model, Eq.\ (\ref{eq:B.7a}). 

In contrast to the $\delta\sigma$-loop expansion above, this
second method does not require knowledge of the
$\delta\sigma$ propagator. It also involves only solving an algebraic equation, no
integrals need to be performed. In contrast to the standard derivation of the 
nonlinear sigma model, which uses Eq.\ (\ref{eq:B.8}), it reqires no explicit knowledge of the
symmetry group and its representations. If no closed-form solution of the saddle-point
equation could be found, one could still solve Eq.\ (\ref{eq:B.11a}) iteratively, since
it is inhomogeneous. In this way one obviously can construct the effective action
perturbatively, order by order in powers of $\pi$. For an analysis of the effective
action to any given order in a loop expansion this would be sufficient. This is the
procedure we follow for the matrix field theory in Sec.\ \ref{subsec:IV.C}.

Finally, we need to worry about the change of the integration measure mentioned
after Eqs.\ (\ref{eqs:B.7}). In the present formalism, this is generated by the
Gaussian fluctuations about $\delta\sigma_{\text{sp}}$. If we write
$\delta\sigma({\bm x}) = \delta\sigma_{\text{sp}}({\bm x}) + \tau({\bm x})$
and integrate out $\tau({\bm x})$ in Gaussian approximation we obtain an
additional contribution to the effective action. The complete effective action  then is
\bea
{\cal A}_{\text{eff}} &=& \frac{1}{2G} \int d{\bm x}\,\left[({\bm\nabla}\pi)^2 
   +({\bm\nabla}\sqrt{1-\pi^2})^2\right] 
\nonumber\\
&& + \Tr\ln\sqrt{1-\pi^2} + O(G),
\label{eq:B.12}
\eea
with $G = 1/c\phi_0^2$ the coupling constant. This is the nonlinear sigma model
including the measure terms, see Eqs.\ (\ref{eqs:B.7}). 


\begin{thebibliography}{66}
\bibitem[{adj()}]{adjoint_footnote}
\bibinfo{note}{The such defined adjoint has the properties $(\eta^+)^+ = \eta$,
  and $(z\eta)^+ = z\eta^+$ for any bispinor $\eta$ and any complex number $z$.
  It is unrelated to the adjoint operation for the fermion fields, which obeys
  $\overline{\bar\psi} = \psi$ and $\overline{z\psi} = z^*{\bar\psi}$. Our
  notation is the same as in Ref.\ \onlinecite{Belitz_Kirkpatrick_1997}.}
\bibitem[{con()}]{contour_footnote}
\bibinfo{note}{Note that the contour for the ${\tilde\Lambda}$ integration must
  start at $-i\infty$ and end at $i\infty$, as can be seen from Eq.\
  (\ref{eq:2.11}). The sign of the ${\bar\Lambda}$ propagator is thus
  consistent with a stable field theory.}
\bibitem[{pow()}]{powers_of_q_footnote}
\bibinfo{note}{Here and in what follows we denote by $O(q^n)$ terms that are of
  $n^{\text{th}}$ order in $q$, $\lambda$, or any combination thereof.}
\bibitem[{fre()}]{frequency_restrictions_footnote}
\bibinfo{note}{Note that, for instance, the first term in Eq.\ (\ref{eq:4.32c})
  can {\em not} be simplified to read $G({\hat\gamma}q +
  {\hat\gamma}q^+)G(\qslash G^{-1}\qslash^+ + \qslash^+ G^{-1} \qslash)$, lest
  the frequency restrictions inherent in $\Phi$ get lost.}
\end{thebibliography}

\end{document}